\documentstyle[12pt]{book} 
\begin{document}

\sloppy
\raggedbottom

\begin{titlepage}

\hfill\vspace{1in}\\
{\Huge\bf\hspace{-\parindent}The Limits of \vspace{3mm}\\
Mathematics--- \vspace{3mm} \\
\rm Course Outline \& \vspace{3mm} \\
Software \vspace{1in} \\
}
{\Large\it To Fran\c{c}oise \vspace{1in} \\
\bf G J Chaitin \vspace{1mm} \\
IBM, P O Box 704 \\
Yorktown Heights, NY 10598 \vspace{1mm} \\
{\it chaitin@watson.ibm.com} \vspace{1in} \\
December 12, 1993
}

\end{titlepage}

\begin{titlepage}

\hfill\\

\end{titlepage}

\markboth
{The Limits of Mathematics}{Course Outline \& Software}

\chapter*{Preface}

In a remarkable development, I have constructed a new definition for a
self-delimiting universal Turing machine (UTM) that is easy to program
and runs very quickly.  This provides a new foundation for algorithmic
information theory (AIT), which is the theory of the size in bits of
programs for self-delimiting UTM's.  Previously, AIT had an abstract
mathematical quality.  Now it is possible to write down executable
programs that embody the constructions in the proofs of theorems.  So
AIT goes from dealing with remote idealized mythical objects to being
a theory about practical down-to-earth gadgets that one can actually
play with and use.

This new self-delimiting UTM is implemented via software written in
Mathematica and C that is based on the related and now largely
obsolete software that I presented in my report ``Exhibiting
randomness in arithmetic using Mathematica and C,'' IBM Research
Report RC-18946, 94 pp., June 1993.  In its turn this report was a
reworking of the software for my book {\it Algorithmic Information
Theory,} Cambridge University Press, 1987.

Using this new software, as well as the latest theoretical ideas, it
is now possible to give a self-contained ``hands on'' mini-course
presenting very concretely my latest proofs of my two fundamental
information-theoretic incompleteness theorems.  The first of these
theorems states that an $N$-bit formal axiomatic system cannot enable
one to exhibit any specific object with program-size complexity
greater than $N+c$.  The second of these theorems states that an
$N$-bit formal axiomatic system cannot enable one to determine more
than $N+c'$ scattered bits of the halting probability $\Omega$.

Most people believe that anything that is true is true for a reason.
These theorems show that some things are true for no reason at all,
i.e., accidentally, or at random.

The latest and I believe the deepest proofs of these two theorems were
originally presented in my paper ``Information-theoretic
incompleteness,'' {\it Applied Mathematics and Computation\/} 52
(1992), pp.\ 83--101.  This paper is reprinted in my book {\it
Information-Theoretic Incompleteness,} World Scientific, 1992.

As is shown in this course, the algorithms considered in the
proofs of these two theorems are now easy to program and run, and by
looking at the size in bits of these programs one can actually, for
the first time, determine exact values for the constants $c$ and $c'$.

It is my intention to use this approach and software in an intensive
short course on the limits of mathematics that I will give at the
University of Maine in Orono and at the University of Copenhagen, in
both cases in the summer of 1994.  I also intend to write up this
course as a new book.  This research report, however, is intended to
make these important new ideas and software widely available in
preliminary form as soon as possible.

I wish to thank Prof.\ George Markowsky of the University of Maine at
Orono and Prof.\ Peter Johansen of the University of Copenhagen for
their stimulating invitations to give intensive summer workshops.  I
also thank Prof.\ John Casti of the Santa Fe Institute, Prof.\ Carlton
Caves of the University of New Mexico, and Prof.\ Cristian Calude of
the University of Auckland for stimulating discussions.

I am grateful to IBM for its enthusiastic and unwavering support of my
research for a quarter of a century, and to my current management
chain at the IBM Research Division, Marty Hopkins, Eric Kronstadt, and
Jeff Jaffe.  Finally I thank the RISC project group, of which I am a
member, for designing the marvelous IBM RISC System/6000 workstations
that I have used for all these calculations, and for providing me with
the latest models of this spectacular computing equipment.

All enquires, comments and suggestions regarding this software should
be sent via e-mail to {\tt chaitin} at {\tt watson.ibm.com}.

\tableofcontents

\newcommand
{\chap}[1]{\chapter*{#1}\markboth{The Limits of Mathematics}{#1}
\addcontentsline{toc}{chapter}{#1}}
\newcommand
{\size}{\small}   

\part{Explanation}

\chap{The New Idea}

First of all, note that we use here the toy LISP from my monograph
{\it Algorithmic Information Theory,} Cambridge University Press,
1987.  (Reprinted with minor revisions thrice, lastly in 1992.)

Here is a quick summary of this toy LISP.  Each LISP primitive
function and variable is as before a single character, but they have
been changed from the IBM APL2 character set to the standard ASCII
character set.  These primitive functions, all of which have a fixed
number of arguments, are now {\tt '} for QUOTE (1 argument), {\tt .}
for ATOM (1 argument), {\tt =} for EQ (2 arguments), {\tt +} for CAR
(1 argument), {\tt -} for CDR (1 argument), {\tt *} for CONS (2
arguments), {\tt \&} for LAMBDA and DEFINE (2 arguments), {\tt :} for
LET-BE-IN (3 arguments), {\tt /} for IF-THEN-ELSE (3 arguments), {\tt
,} for OUTPUT (1 argument), {\tt !} for EVAL (1 argument), and {\tt ?}
for depth-limited EVAL (had 2 arguments, now has 3).  The
meta-notation {\tt "} indicates that an S-expression with explicit
parentheses follows, not what is usually the case in my toy LISP, an
M-expression, in which the parentheses for each primitive function are
implicit.  Finally the empty list NIL is written {\tt ()}, and TRUE
and FALSE are {\tt 1} and {\tt 0}.

The new idea that goes beyond what is presented in my Cambridge
University Press monograph is this.  We define our standard
self-delimiting universal Turing machine as follows.  Its program is
in binary, and appears on a tape in the following form.  First comes a
toy LISP expression, written in ASCII with 7 bits per character.  Note
that this expression is self-delimiting because parentheses must
balance.  The TM reads in this LISP expression, and then evaluates it.
As it does this, two new primitive functions {\tt @} and {\tt \%} with
no arguments may be used to read more from the TM tape.  Both of these
functions explode if the tape is exhausted, killing the computation.
{\tt @} reads a single bit from the tape, and {\tt \%} reads in an
entire LISP expression, in 7-bit character chunks.

This is the only way that information on the TM tape may be accessed,
which forces it to be used in a self-delimiting fashion.  This is
because no algorithm can search for the end of the tape and then use
the length of the tape as data in the computation.  If an algorithm
attempts to read a bit that is not on the tape, the algorithm aborts.

How is information placed on the TM tape in the first place?  Well, in
the starting environment, the tape is empty and any attempt to read it
will give an error message.  To place information on the tape, a new
argument has been added to the primitive function {\tt ?} for
depth-limited evaluation.

Consider the three arguments $\alpha$, $\beta$ and $\gamma$ of {\tt
?}.  The meaning of the first argument, the depth limit $\alpha$, has
been changed slightly.  If $\alpha$ is a non-null atom, then there is
no depth limit.  If $\alpha$ is the empty list, this means zero depth
limit (no function calls or re-evaluations).  And an $N$-element list
$\alpha$ means depth limit $N$.  The second argument $\beta$ of {\tt
?} is as before the expression to be evaluated as long as the depth
limit $\alpha$ is not exceeded.  The new third argument $\gamma$ of
{\tt ?} is a list of bits to be used as the TM tape.

The value $\nu$ returned by the primitive function {\tt ?} is also
changed.  $\nu$ is a list.  The first element of $\nu$ is {\tt !} if
the evaluation of $\beta$ aborted because an attempt was made to read
a non-existent bit from the TM tape.  The first element of $\nu$ is
{\tt ?} if evaluation of $\beta$ aborted because the depth limit
$\alpha$ was exceeded.  {\tt !} and {\tt ?} are the only possible
error flags, because my toy LISP is designed with maximally permissive
semantics.  If the computation $\beta$ terminated normally instead of
aborting, the first element of $\nu$ will be a list with only one
element, which is the result produced by the computation $\beta$.
That's the first element of the list $\nu$ produced by the {\tt ?}
primitive function.

The rest of the value $\nu$ is a stack of all the arguments to the
primitive function {\tt ,} that were encountered during the evaluation
of $\beta$.  More precisely, if {\tt ,} was called $N$ times during
the evaluation of $\beta$, then $\nu$ will be a list of $N+1$
elements.  The $N$ arguments of {\tt ,} appear in $\nu$ in inverse
chronological order.  Thus {\tt ?} can not only be used to determine
if a computation $\beta$ reads too much tape or goes on too long
(i.e., to greater depth than $\alpha$), but {\tt ?} can also be used
to capture all the output that $\beta$ displayed as it went along,
whether the computation $\beta$ aborted or not.

In summary, all that one has to do to simulate a self-delimiting
universal Turing machine $U(p)$ running on the binary program $p$ is
to write
\begin{verbatim}
                         ?0'!%p
\end{verbatim}
This is an M-expression with parentheses omitted from primitive
functions, because all primitive functions have a fixed number of
arguments.  With all parentheses supplied, it becomes the S-expression
\begin{verbatim}
                     (?0('(!(%)))p)
\end{verbatim}
This says that one is to read a complete LISP S-expression from the TM
tape $p$ and then evaluate it without any time limit and using
whatever is left on the tape $p$.

Two more primitive functions have also been added, the 2-argument
function \verb|^| for APPEND, i.e., list concatenation, and the
1-argument function {\tt \#} for converting an expression into the
list of the bits in its ASCII character string representation.  These
are used for constructing the bit strings that are then put on the TM
tape using {\tt ?}'s third argument $\gamma$.  Note that the functions
\verb|^|, {\tt \#} and {\tt \%} could be programmed rather than
included as built-in primitive functions, but it is extremely
convenient and much much faster to provide them built in.

Finally a new 1-argument identity function \verb|~| with the
side-effect of outputting its argument is provided for debugging.
Output produced by \verb|~| is invisible to the ``official'' {\tt ?}
and {\tt ,} output mechanism.  \verb|~| is needed because {\tt
?}$\alpha\beta\gamma$ suppresses all output $\theta$ produced within
its depth-controlled evaluation of $\beta$.  Instead {\tt ?} stacks
all output $\theta$ from within $\beta$ for inclusion in the final
value $\nu$ that {\tt ?} returns, namely $\nu = $ (atomic error flag
or (value of $\beta$) followed by the output $\theta$).

\chap{Course Outline}

The course begins by explaining with examples my toy LISP.  Possibly
the theory of LISP program-size complexity is developed a little,
following my recent papers ``LISP program-size complexity,'' {\it
Applied Mathematics and Computation\/} 49 (1992), pp.\ 79--93, ``LISP
program-size complexity II,'' {\it Applied Mathematics and
Computation\/} 52 (1992), pp.\ 103--126, ``LISP program-size
complexity III,'' {\it Applied Mathematics and Computation\/} 52
(1992), pp.\ 127--139, ``LISP program-size complexity IV,'' {\it
Applied Mathematics and Computation\/} 52 (1992), pp.\ 141--147.
These papers are reprinted in my book {\it Information-Theoretic
Incompleteness,} World Scientific, 1992.

LISP program-size complexity is extremely simple and concrete.  In
particular, it is easy to show that it is impossible to prove that a
self-contained LISP expression is elegant, i.e., that no smaller
expression has the same value.  To prove that an $N$-character LISP
expression is elegant requires a formal axiomatic system that itself
has LISP complexity $N$.  Also, LISP program-size complexity is
subadditive, because expressions are self-delimiting and can be
concatenated, and also because we are dealing with pure LISP and no
side-effects get in the way.  Moreover, the probability $\Omega_{\rm
LISP}$ that a LISP expression ``halts'' or has a value is
well-defined, also because programs are self-delimiting.  Finally, it
is easy to see that the LISP program-size complexity of the first $N$
bits of the LISP halting probability $\Omega_{\rm LISP}$ grows
linearly with $N$.  Therefore to be able to determine the first $N$
bits of $\Omega_{\rm LISP}$ requires a formal axiomatic system whose
LISP complexity also grows linearly with $N$.

It should then be pointed out that LISP programs have severe
information-theoretic limitations because they do not encode
information very efficiently in 7-bit ASCII characters subject to LISP
syntax constraints.  Arbitrary binary programs are denser and much
better, but they should at least be kept self-delimiting.

So next we define our standard self-delimiting universal Turing
machine $U(p)$ using
\begin{verbatim}
                         ?0'!%p
\end{verbatim}
as explained in the previous chapter.

Next we show that
\[
   H(x,y) \le H(x) + H(y) + c.
\]
Here $H(\cdots)$ denotes the size in bits of the smallest program that
makes our standard universal Turing machine compute $\cdots$.  Thus
this inequality states that the information needed to compute the pair
$(x,y)$ is bounded by a constant $c$ plus the sum of the information
needed to compute $x$ and the information needed to compute $y$.
Consider
\begin{verbatim}
                        *!%*!%()
\end{verbatim}
This is an M-expression with parentheses omitted from primitive
functions.  With all parentheses supplied, it becomes the S-expression
\begin{verbatim}
                  (*(!(%))(*(!(%))()))
\end{verbatim}
In fact, $c$ is just 7 times the size in characters of this LISP
S-expression, which is exactly 20 characters.  Thus $c = 7 \times 20 =
140$ bits!  See {\tt univ.lisp}.  Note that in standard LISP this
would be something like
\begin{verbatim}
              (CONS (EVAL (READ-EXPRESSION))
              (CONS (EVAL (READ-EXPRESSION))
                    NIL))
\end{verbatim}
which is much more than 20 characters long.

Looking at binary strings $x$ with size $|x|$ in bits, we next show
that
\[
   H(x) \le 2|x| + c
\]
and
\[
   H(x) \le |x| + H(|x|) + c'.
\]
As before, the programs for doing this are exhibited and run.
Also $c$ and $c'$ are determined.
See {\tt univ.lisp}.

Next we show how to calculate the halting probability $\Omega$ of our
standard self-delimiting universal Turing machine in the limit from
below.  The LISP program for doing this, {\tt omega.lisp}, is now
remarkably clear and fast, and much better than the one given in my
Cambridge University Press monograph.  (See also {\tt omega2.lisp} and
{\tt omega3.lisp}.)  Using this, we show that if $\Omega_N$ denotes
the first $N$ bits of the fractional part of the base-two real number
$\Omega$, then
\[
   H(\Omega_N) \ge N - c.
\]
Again this is done with a program that can actually be run and whose
size gives us a value for $c$.  See {\tt omega4.lisp}.

Next we turn to the self-delimiting program-size complexity $H(X)$ for
infinite r.e.\ sets $X$, which is not considered at all in my
Cambridge University Press monograph.  This is now defined to be the
size in bits of the smallest LISP expression $\xi$ that executes
forever without halting and outputs the members of the r.e.\ set $X$
using the LISP primitive {\tt ,}.  {\tt ,} is an identity function
with the side-effect of outputting the value of its argument.  Note
that this LISP expression $\xi$ is allowed to read additional bits or
expressions from the TM tape using the primitive functions {\tt @} and
{\tt \%} if $\xi$ so desires.  But of course $\xi$ is charged for
this; this adds to $\xi$'s program size.

It was in order to deal with such unending expressions $\xi$ that the
LISP primitive function for time-limited evaluation {\tt ?} now
captures all output from {\tt ,} within its second argument $\beta$.

To illustrate these new concepts, we show that
\[
   H(X \cap Y) \le H(X) + H(Y) + c
\]
and
\[
   H(X \cup Y) \le H(X) + H(Y) + c'
\]
for infinite r.e.\ sets $X$ and $Y$.  As before, we run examples and
determine $c$ and $c'$.  See {\tt sets0.lisp} through {\tt
sets4.lisp}.

Now consider a formal axiomatic system $A$ of complexity $N$,
i.e., with a set of theorems $T_A$ that considered as an r.e. set as
above has self-delimiting program-size complexity $H(T_A)$.  We show
that $A$ has the following limitations.  First of all, we show
directly that $A$ cannot enable us to exhibit a specific S-expression
$s$ with self-delimiting complexity $H(s)$ greater than $N+c$.  See
{\tt godel.lisp} and {\tt godel2.lisp}.

Secondly, using the lower bound of $N-c$ on $H(\Omega_N)$ established
in {\tt omega4.lisp}, we show that $A$ cannot enable us to determine
not only the first $N+c'$ bits of $\Omega$, but any $N+c'$ bits of
$\Omega$, even if they are scattered and we leave gaps.  (See {\tt
godel3.lisp}.)  In my Cambridge University Press monograph, this took
a hundred pages to show, and involved the systematic development of a
general theory using measure-theoretic arguments.  Following
``Information-theoretic incompleteness,'' {\it Applied Mathematics and
Computation\/} 52 (1992), pp.\ 83--101, the proof is now a
straight-forward Berry-paradox-like program-size argument.  Moreover
we are using a deeper definition of $H(A) \equiv H(T_A)$ via infinite
self-delimiting computations.

Finally, determining the bits of $\Omega$ is dressed up as a question
involving the number of solutions of a very large diophantine
equation.  How to do this is indicated at the end of the next chapter.
See {\tt lisp.rm} and {\tt eq.m}.

And last but not least, the philosophical implications of all this
should be discussed, especially the extent to which it tends to
justify experimental mathematics.  This would be along the lines of
the discussion in my talk transcript ``Randomness in arithmetic and
the decline and fall of reductionism in pure mathematics,'' {\it
Bulletin of the European Association for Theoretical Computer
Science,} No.\ 50 (June 1993), pp.\ 314--328.

This concludes our ``hand-on'' mini-course on the
information-theoretic limits of mathematics.

\chap{Software User Guide}

All the software for this course is written in {\sl Mathematica}.
This {\sl Mathematica} code is remarkably compact, but it is sometimes
slow.  So one {\sl C} program plus equipment for automatically
generating another is also provided.

I used Version 2.1 of {\sl Mathematica} as described in the second
edition of Wolfram's book {\it Mathematica---A System for Doing
Mathematics by Computer,} running on {\sl IBM RISC} System/6000
workstations.

There are seven different kinds of files:

\begin{itemize}
\item Included in this distribution:

\begin{enumerate}
\item {\tt *.m} files are {\sl Mathematica} code.
\item {\tt *.c} files are {\sl C} code.
\item {\tt *.lisp} files are toy {\sl LISP} code.
\item {\tt *.rm} are register machine code.
\end{enumerate}

\item These will produce:

\begin{itemize}
\item[\rm 5.] {\tt *.xrm} files are expanded register machine code
      (lower level code than that in {\tt *.rm} files).
\item[\rm 6.] {\tt *.run,~*.srun,~*.mrun,~*.crun,~*.xcrun,~*.cmrun} files
      are the output from {\sl LISP} interpreter runs.
\item[\rm 7.] {\tt *.eq} files are exponential diophantine equations.
\end{itemize}

\end{itemize}

Six different {\sl LISP} interpreters are included here:

\begin{enumerate}

\item
{\tt lisp.m} is a {\sl LISP} interpreter written in {\sl
Mathematica} that uses {\sl Mathematica} list structures to
represent {\sl LISP} S-expressions.  Bindings are kept in a fast
look-up table.  {\tt lisp.m} converts an {\tt X.lisp} input file into
an {\tt X.run} output file.

\[
\mbox{\tt X.lisp}
\longrightarrow \fbox{\tt lisp.m} \longrightarrow
\mbox{\tt X.run}
\]

\item
{\tt slisp.m} is a {\sl LISP} interpreter written in {\sl
Mathematica} that uses {\sl Mathematica} character strings to
represent {\sl LISP} S-expressions.  Bindings are kept in an
association list that must be searched sequentially.  {\tt slisp.m}
converts an {\tt X.lisp} input file into an {\tt X.srun} output file.

\[
\mbox{\tt X.lisp}
\longrightarrow \fbox{\tt slisp.m} \longrightarrow
\mbox{\tt X.srun}
\]

\item
{\tt lispm.m} is a {\sl Mathematica} program that simulates a {\sl
LISP} interpreter running on a register machine.  {\tt lispm.m}
converts an {\tt X.lisp} input file into an {\tt X.mrun} output file.

Before running this program, {\tt xpnd.m} must be used to convert
{\tt lisp.rm} into {\tt lisp.xrm}, which is needed by this program.

\[
\mbox{\tt X.lisp}
\longrightarrow
\begin{array}[b]{c}
\mbox{\tt lisp.rm} \\
\downarrow         \\
\fbox{\tt xpnd.m}  \\
\downarrow         \\
\mbox{\tt lisp.xrm}\\
\downarrow         \\
\fbox{\tt lispm.m}
\end{array}
\longrightarrow
\mbox{\tt X.mrun}
\]

\item
{\tt clisp.m} is a {\sl Mathematica} program serving as a driver for a
{\sl LISP} interpreter written in {\sl C}.  {\tt clisp.m} converts an
{\tt X.lisp} input file into an {\tt X.crun} output file.

Before running {\tt clisp.m}, the {\sl C} program {\tt lisp.c} must be
compiled.  In {\sl UNIX} this is done using the command {\tt cc -O
-olisp lisp.c}.

As written, {\tt clisp.m} will only work properly running on {\sl
UNIX}.  For other styles of operating system, either {\tt clisp.m}
must be modified, or {\tt lisp.m} should be used instead.

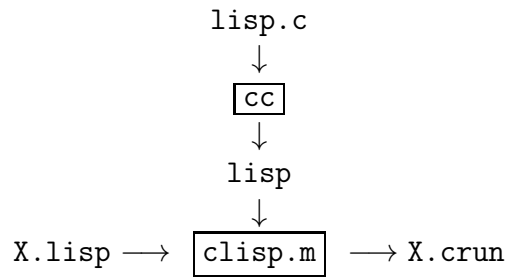
\[
\mbox{\tt X.lisp}
\longrightarrow
\begin{array}[b]{c}
\mbox{\tt lisp.c}\\
\downarrow       \\
\fbox{\tt cc}    \\
\downarrow       \\
\mbox{\tt lisp}  \\
\downarrow       \\
\fbox{\tt clisp.m}
\end{array}
\longrightarrow
\mbox{\tt X.crun}
\]

\item
{\tt xclisp.m} is a {\sl Mathematica} program serving as a driver for a
{\sl LISP} interpreter written in {\sl C}.  {\tt xclisp.m} converts an
{\tt X.lisp} input file into an {\tt X.xcrun} output file.

Before running {\tt xclisp.m}, the {\sl C} program {\tt lisp.c} must be
compiled.  In {\sl UNIX} this is done using the command {\tt cc -O
-olisp lisp.c}.

As written, {\tt xclisp.m} will only work properly running on {\sl
UNIX}.  For other styles of operating system, either {\tt xclisp.m}
must be modified, or {\tt lisp.m} should be used instead.

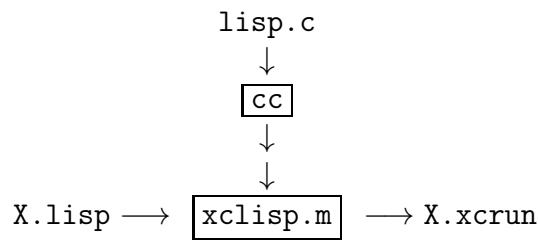
\[
\mbox{\tt X.lisp}
\longrightarrow
\begin{array}[b]{c}
\mbox{\tt lisp.c}\\
\downarrow       \\
\fbox{\tt cc}    \\
\downarrow       \\

\downarrow       \\
\fbox{\tt xclisp.m}
\end{array}
\longrightarrow
\mbox{\tt X.xcrun}
\]

\item
{\tt clispm.m} is a {\sl Mathematica} program serving as a driver for
a {\sl C} program that simulates a {\sl LISP} interpreter running on a
register machine.  {\tt clispm.m} converts an {\tt X.lisp} input file
into an {\tt X.cmrun} output file.

Before running {\tt clispm.m}, {\tt xpnd.m} must be used to convert
{\tt lisp.rm} into {\tt lisp.xrm}.  {\tt rm2c.m} must then be used to
convert {\tt lisp.xrm} into the {\sl C} program {\tt lispm.c}.  {\tt
lispm.c} is then compiled.  In {\sl UNIX} this is done using the
command {\tt cc -O -olispm lispm.c}.

As written, {\tt clispm.m} and {\tt rm2c.m} will only work properly
running on {\sl UNIX}.  For other styles of operating system, either
{\tt clispm.m} and {\tt rm2c.m} must be modified, or {\tt lispm.m}
should be used instead.

\[
\mbox{\tt X.lisp}
\longrightarrow
\begin{array}[b]{c}
\mbox{\tt lisp.rm} \\
\downarrow         \\
\fbox{\tt xpnd.m}  \\
\downarrow         \\
\mbox{\tt lisp.xrm}\\
\downarrow         \\
\fbox{\tt rm2c.m}  \\
\downarrow         \\
\mbox{\tt lispm.c} \\
\downarrow         \\
\fbox{\tt cc}      \\
\downarrow         \\
\mbox{\tt lispm}   \\
\downarrow         \\
\fbox{\tt clispm.m}
\end{array}
\longrightarrow
\mbox{\tt X.cmrun}
\]

\end{enumerate}

To run any one {\tt X.m} of these six {\sl LISP} interpreters, first
enter {\sl Mathematica} using the command {\tt math}.  Then tell
{\sl Mathematica},
\[
\mbox{\tt << X.m}
\]
To run a {\sl LISP} program {\tt X.lisp}, enter
\[
\mbox{\tt run @ "X"}
\]
To run several programs, enter
\begin{center}
\verb!run /@ {"X","Y","Z"}!
\end{center}
Before changing to another {\sl LISP} interpreter, type {\tt Exit} to
exit from {\sl Mathematica}, and then begin a fresh {\sl Mathematica}
session.

Here is how to run the programs that compute the halting probability
$\Omega$ in the limit from below:

\begin{verbatim}
               math
               << clisp.m
               run /@ {"omega","omega2","omega3"}
               run @ "omega4"
               Exit
\end{verbatim}

The six different {\sl LISP} interpreters run at vastly different
speeds, but should always produce identical results.  This can easily
be checked, for example in {\sl UNIX} as follows:

\begin{verbatim}
               diff X.run X.crun > out
               vi out
\end{verbatim}

These six {\sl LISP} interpreters all use the same frontend, {\tt
frontend.m.} {\tt frontend.m} is written in {\sl Mathematica}.  As
each M-expression is read in, it is written out, then converted to an
S-expression that is written out and evaluated.\footnote{The
conversion from M-to S-expression mostly consists of making all
implicit parentheses explicit.}

One register machine program {\tt *.rm} is provided: {\tt lisp.rm}.
{\tt lisp.rm} is the {\sl LISP} interpreter used by {\tt lispm.m} and
{\tt clispm.m}, and converted into the monster exponential diophantine
equation by {\tt eq.m}.

More precisely, to convert the register machine
program {\tt X.rm} into an exponential diophantine equation there are
two steps.  First, use {xpnd.m} to convert {\tt X.rm} into {\tt
X.xrm}.  Then use {\tt eq.m} to convert {\tt X.xrm} into {\tt X.eq}.
For more output, set {\tt fulloutput = True} before typing {\tt <<
eq.m}.  For each conversion, a fresh copy of {\tt eq.m} must be loaded
into a clean {\sl Mathematica} session.

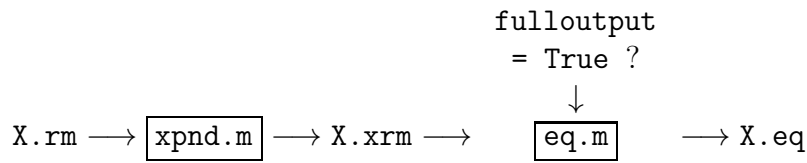
\[
\mbox{\tt X.rm} \longrightarrow
\fbox{\tt xpnd.m} \longrightarrow
\mbox{\tt X.xrm} \longrightarrow
\begin{array}[b]{c}
\mbox{\tt fulloutput} \\
\mbox{\tt = True \rm ?} \\
\downarrow \\
\fbox{\tt eq.m}
\end{array}
\longrightarrow
\mbox{\tt X.eq}
\]

Here is how to generate the monster equation:

\begin{verbatim}
               math
               << xpnd.m
               run @ "lisp"
               Exit

               math
              [fulloutput = True]
               << eq.m
               fn of fn.xrm file = lisp
               Exit
\end{verbatim}

This software enables us to dress $\Omega$ up as a diophantine
equation as follows.

Take the equation in {\tt lisp.eq}.  Substitute a toy {\tt LISP}
expression that halts if and only if (the $k$th bit of the $n$th
approximation to $\Omega$ is 1) for {\tt input[reg\$expression]}.
(Most of the pieces for this are in {\tt omega.lisp}.)  Substitute
the appropriate large constants for the ASCII look-up tables
\begin{verbatim}
               input[reg$ascii$bits]
               input[reg$all$ascii$characters]
               input[reg$printable$ascii$characters]
\end{verbatim}
(See {\tt lispm.m} and {\tt rm2c.m} and the {\tt s2i} function in {\tt
eq.m} for the details of how to construct these very large constants.)
Substitute 0 for {\tt input[reg\$X]} for each other register {\tt
reg\$X}.

The resulting exponential diophantine equation is $1.7 \times 10^6$
characters long and has $2.6 \times 10^4$ variables.  It has exactly
one solution for a given value of $k$ and $n$ if the $k$th bit of the
$n$th approximation to $\Omega$ is 1.  It has no solutions for a given
value of $k$ and $n$ if the $k$th bit of the $n$th approximation to
$\Omega$ is 0.  Now think of $n$ as a variable rather than as a
parameter.  The resulting equation has only finitely many solutions if
the $k$th bit of $\Omega$ is 0.  It has infinitely many solutions if
the $k$th bit of $\Omega$ is 1.

\chap{Bibliography}

\begin{itemize}
\item[{[1]}] S. Wolfram, {\it Mathematica---A System for Doing
Mathematics by Computer,} second edition, Addison-Wesley, 1991.
\item[{[2]}] B. W. Kernighan and D. M. Ritchie, {\it The C Programming
Language,} second edition, Prentice Hall, 1988.
\item[{[3]}] G. J. Chaitin, {\it Algorithmic Information Theory,}
revised third printing, Cambridge University Press, 1990.
\item[{[4]}] G. J. Chaitin, {\it Information, Randomness \&
Incompleteness,} second edition, World Scientific, 1990.
\item[{[5]}] G. J. Chaitin, ``LISP program-size complexity,'' {\it
Applied Mathematics and Computation\/} 49 (1992), pp.\ 79--93.
\item[{[6]}] G. J. Chaitin, ``Information-theoretic incompleteness,''
{\it Applied Mathematics and Computation\/} 52 (1992), pp.\ 83--101.
\item[{[7]}] G. J. Chaitin, ``LISP program-size complexity II,'' {\it
Applied Mathematics and Computation\/} 52 (1992), pp.\ 103--126.
\item[{[8]}] G. J. Chaitin, ``LISP program-size complexity III,'' {\it
Applied Mathematics and Computation\/} 52 (1992), pp.\ 127--139.
\item[{[9]}] G. J. Chaitin, ``LISP program-size complexity IV,'' {\it
Applied Mathematics and Computation\/} 52 (1992), pp.\ 141--147.
\item[{[10]}] G. J. Chaitin, {\it Information-Theoretic
Incompleteness,} World Scientific, 1992.
\item[{[11]}] G. J. Chaitin, ``Randomness in arithmetic and the
decline and fall of reductionism in pure mathematics,'' {\it Bulletin
of the European Association for Theoretical Computer Science,} No.\ 50
(June 1993), pp.\ 314--328.
\item[{[12]}] G. J. Chaitin, ``Exhibiting randomness in arithmetic
using Mathematica and C,'' IBM Research Report RC-18946, June 1993.
\item[{[13]}] G. J. Chaitin, ``On the number of $n$-bit strings with
maximum complexity,'' {\it Applied Mathematics and Computation\/} 59
(1993), pp.\ 97--100.
\item[{[14]}] G. J. Chaitin, ``Randomness in arithmetic and the limits
of mathematical reasoning,'' in J. Tr\^an Thanh V\^an, {\it
Proceedings of the Vth Rencontres de Blois}, Editions Fronti\`eres, in
press.
\item[{[15]}] C. Calude, {\it Information and Randomness,}
Springer-Verlag, in press.
\item[{[16]}] K. Svozil, {\it Randomness \& Undecidability in
Physics,} World Scientific, in press.
\end{itemize}

\part{The Course}

{
}\chap{univ.lisp}{\size\begin{verbatim}
[[[
 First steps with my new construction for
 a self-delimiting universal Turing machine.
 We show that
    H(x,y) <= H(x) + H(y) + c
 and determine c.
 Consider a bit string x of length |x|.
 We also show that
    H(x) <= 2|x| + c
 and that
    H(x) <= |x| + H(|x|) + c
 and determine both these c's.
]]]

[first demo the new lisp primitive functions]
^'(1234567890)'(abcdefghi)
@
?0 '@ '()
?0 '@ '(1)
?0 '@ '(0)
?0 '@ '(x)
?0 '**@()**@()() '(10)
?0 '**,@()**,@()() '(10)
?0 '**,@()**,@()**,@()() '(10)
#'a
#'(abcdef)
#'(12(34)56)
?0 '% '(110 0001)
?0 '% '(110 0010)
?0 '% '(110 0011)
?0 '% '(110 0100)
?0 '% '(110 0101)
?0 '% '(110 0110)
?0 '% '(110 0111)
?0 '% #'a
?0 '% '(010 100)
?0 '% '(010 1001)
?0 '% '(010 1000 011 0001 011 0010 010 1001)
?0 '% '(010 1000 011 0001 011 0010         )
?0 '% '(010 1000 011 0001 011 001          )
?0 '% '(010 100                            )
?0 '% #'(abcdef)
?0 '% #'(12(34)56)
?0 '*%*%() ,^ #'a #'b
?0
':(f) :x@ :y@ /=xy *x(f) () (f)
'(0011001101)
#':(f) :x@ :y@ /=xy *x(f) () (f)
[ Here is the self-delimiting universal Turing machine! ]
[ (with slightly funny handling of out-of-tape condition) ]
& (Up) ++?0'!%p
[Show that H(x) <= 2|x| + c]
(U
 ^ ,#,':(f) :x@ :y@ /=xy *x(f) () (f)
   '(0011001101)
)
(U
 ^ ,#,':(f) :x@ :y@ /=xy *x(f) () (f)
   '(0011001100)
)
(U
 ^ ,#,'*!%*!%() [The length of this bit string is the]
                [constant c in H(x,y) <= H(x)+H(y)+c.]
 ^ #':(f) :x@ :y@ /=xy *x(f) () (f)
 ^ '(0011001101)
 ^ #':(f) :x@ :y@ /=xy *x(f) () (f)
   '(1100110001)
)
[Size of list in reverse decimal!]
& (Se) /.e() (I[,](S-e))
[Increment reverse decimal]
& (In) /.n'(1) :d+n :r-n
     /=d0*1r
     /=d1*2r
     /=d2*3r
     /=d3*4r
     /=d4*5r
     /=d5*6r
     /=d6*7r
     /=d7*8r
     /=d8*9r
    [/=d9]*0(Ir)
[Reverse list]
& (Re) /.e() ^(R-e)*+e()
[Convert to binary and display size in decimal]
& (Me) :e [,]#[,]e :f ,(R[,](Se)) e
(M'a)
(M'())
& (Dk) /=0+k *1(D-k) /.-k () *0-k [D = decrement]
,(D,(D,(D,(D,'(001)))))
(U
 ^ ,#,'   [Show that H(x) <= |x| + H(|x|) + c]
   : (Re) /.e() ^(R-e)*+e()          [R = reverse  ]
   : (Dk) /=0+k *1(D-k) /.-k () *0-k [D = decrement]
   : (Lk) /.k () *@(L(Dk))           [L = loop     ]
   (L(R!%))
 ^ #''(1000)
   '(0000 0001)
)
\end{verbatim}
}\chap{omega.lisp}{\size\begin{verbatim}

[[[[ Omega in the limit from below! ]]]]

[Look at the behavior of typical 7-bit programs]
?0'!,%'(010 1000) [lpar]
?0'!,%'(010 1001) [rpar]
?0'!,%'(011 0001) [1]
[All strings of length k / with same length as k ]
& (Xk) /.k'(()) (Z(X-k))
[Append 0 and 1 to each element of list]
& (Zx) /.x() *^+x'(0) *^+x'(1) (Z-x)
(Z'((a)(b)))
(X'())
(X'(1))
(X'(11))
(X'(111))
[Size of list in reverse binary]
& (Se) /.e() (I(S-e))
[Increment reverse binary]
& (Ix) /.x'(1) /=+x0 *1-x *0(I-x)
(S'())
(S'(a))
(S'(ab))
(S'(abc))
(S'(abcd))
[Pad x to length of k on right and reverse]
& (Rxk) /.x /.k() *0(Rx-k) ^(R-x-k)*+x()
(R'(1)'(11))
(R'(01)'(1111))
(R'(0001)'(1111 1111))
[Set of programs in x that halt within time k]
& (Hxk) /.x() /,=0.+?k'!%,+x *+x(H-xk) (H-xk)
(H '((111)(111 1111)(000)(000 0000)) '0)
(H , *#'a *#'('(xy)) *#':(X)(X)(X) () '(111))
[For LISP omega must separate read & exec.  ]
& (Gxk) /.x()  [version of H for lisp omega ]
 : e  ?0'%,+x  [read expression from prog +x]
 [If read finished, evaluate exp for time k ]
 [with empty tape, so @ and % will fail!    ]
 : v  /.+e e ?k++e()[run for time k, no tape]
 /,=0.+v  *+x(G-xk) [program +x halted      ]
          (G-xk)    [program +x didn't halt ]
(G '((111)(111 1111)(000)(000 0000)) '0)
(G , *#'a *#'('(xy)) *#':(X)(X)(X) () '(111))
(H , *^#'@'(1) *^#'%#'(ab) () '(111))
(G , *^#'@'(1) *^#'%#'(ab) () '(111))
[Omega sub k!]
& (Wk) *0*".(R,(S,(H,(Xk)k))k)
(W'())
(W'(1))
(W'(11))
(W'(111))
[[[[
(W'(111 1))
(W'(111 11))
(W'(111 111))
]]]]
(W'(111 1111))
[[[[
(W'(111 1111 1))
]]]]
\end{verbatim}
}\chap{omega2.lisp}{\size\begin{verbatim}

[[[[ Omega in the limit from below! ]]]]

[All strings of length k / with same length as k ]
& (Xk) /.k'(()) (Z(X-k))
[Append 0 and 1 to each element of list]
& (Zx) /.x() *^+x'(0) *^+x'(1) (Z-x)
[Size of list in reverse binary]
& (Se) /.e() (I(S-e))
[Increment reverse binary]
& (Ix) /.x'(1) /=+x0 *1-x *0(I-x)
[Pad x to length of k on right and reverse]
& (Rxk) /.x /.k() *0(Rx-k) ^(R-x-k)*+x()
[Set of programs in x that halt within time k]
& (Hxk) /.x() /=0.+?k'!%+x *+x(H-xk) (H-xk)
[Omega sub k!]
& (Wk) *0*".(R,(S,(H,(Xk)k))k)
(W'())
(W'(1))
(W'(11))
(W'(111))
(W'(111 1))
(W'(111 11))
(W'(111 111))
(W'(111 1111))
(W'(111 1111 1))
\end{verbatim}
}\chap{omega3.lisp}{\size\begin{verbatim}

[[[[ Omega in the limit from below! ]]]]

[All strings of length k / with same length as k ]
& (Xk) /.k'(()) (Z(X-k))
[Append 0 and 1 to each element of list]
& (Zx) /.x() *^+x'(0) *^+x'(1) (Z-x)
[Size of list in reverse binary]
& (Se) /.e() (I(S-e))
[Increment reverse binary]
& (Ix) /.x'(1) /=+x0 *1-x *0(I-x)
[Pad x to length of k on right and reverse]
& (Rxk) /.x /.k() *0(Rx-k) ^(R-x-k)*+x()
[Set of programs in x that halt within time k]
& (Hxk) /.x() /=0.+?k'!%+x *+x(H-xk) (H-xk)
[Omega sub k!]
& (Wk) *0*".(R(S(H(Xk)k))k)
(W'())
(W'(1))
(W'(11))
(W'(111))
(W'(111 1))
(W'(111 11))
(W'(111 111))
(W'(111 1111))
(W'(111 1111 1))
(W'(111 1111 11))
(W'(111 1111 111))
(W'(111 1111 111 1))
(W'(111 1111 111 11))
(W'(111 1111 111 111))
(W'(111 1111 111 1111))
[The following exhaust storage:]
[[[[
(W'(111 1111 111 1111 1))
(W'(111 1111 111 1111 11))
]]]]
\end{verbatim}
}\chap{omega4.lisp}{\size\begin{verbatim}

[[[
 Show that H(Omega sub n) > n - c and determine c.
 Omega sub n is the first n bits of Omega.
]]]

[First test new stuff]

[Compare two fractional binary numbers, i.e., is 0.x < 0.y ?]
& (<xy) /.x /.y   0
                  (<'(0)y)
            /.y   0
                  /+x /+y (<-x-y)
                          0
                      /+y 1
                          (<-x-y)
(<'(000)'(000))
(<'(000)'(001))
(<'(001)'(000))
(<'(001)'(001))
(<'(110)'(110))
(<'(110)'(111))
(<'(111)'(110))
(<'(111)'(111))
(<'()'(000))
(<'()'(001))
(<'(000)'())
(<'(001)'())

[Now run it all!]

++?0'!%

^,#,'

[All strings of length k / with same length as k ]
: (Xk) /.k'(()) (Z(X-k))
[Append 0 and 1 to each element of list]
: (Zx) /.x() *^+x'(0) *^+x'(1) (Z-x)
[Size of list in reverse binary]
: (Se) /.e() (I(S-e))
[Increment reverse binary]
: (Ix) /.x'(1) /=+x0 *1-x *0(I-x)
[Pad x to length of k on right and reverse]
: (Rxk) /.x /.k() *0(Rx-k) ^(R-x-k)*+x()
[Set of programs in x that halt within time k]
: (Hxk) /.x() /~=0.+?k'!%~+x *+x(H-xk) (H-xk)
[Omega sub k without 0. at beginning
 (i.e. only fractional part).]
: (Wk) (R(S(H(Xk)k))k)

[Compare two fractional binary numbers, i.e., is 0.x < 0.y ?]
: (<xy) /.x /.y   0
                  (<'(0)y)
            /.y   0
                  /+x /+y (<-x-y)
                          0
                      /+y 1
                          (<-x-y)
: w !%            [Read and execute from remainder of tape
                   a program to compute an n-bit
                   initial piece of Omega.]
: (Lk)            [Main Loop]
  : x    (Wk)     [Compute the kth lower bound on Omega]
  /(<xw) (L*1k)   [Are the first n bits OK?  If not, bump k.]
         (B(Xk))  [Form the union of all output of k-bit
                   programs within time k, output it,
                   and halt.
                   This is bigger than anything of complexity
                   less than or equal to n!]
[This total output will be bigger than each individual output,
 and therefore must come from a program with more than n bits.
]
[Union of all output of programs in list p within time k.]
: (Bp) /.p() * ~?k'!%~+p (B-p) [ k is implicit argument.]

(L())         [Start main loop running with k initially zero.]

,#,'

'(1111)       [These really are the first 4 bits of Omega!]
\end{verbatim}
}\chap{sets0.lisp}{\size\begin{verbatim}
[[[
 Test basic (finite) set functions.
]]]

[Set membership predicate; is e in set s?]
& (Mes) /.s0 /=e+s1 (Me-s)
(Mx'(12345xabcdef))
(Mq'(12345xabcdef))
[Eliminate duplicate elements from set s]
& (Ds) /.s() /(M+s-s) (D-s) *+s(D-s)
(D'(1234512345abcdef))
(D(D'(1234512345abcdef)))
[Set union]
& (Uxy) /.xy /(M+xy) (U-xy) *+x(U-xy)
(U'(12345abcdef)'(abcdefUVWXYZ))
[Set intersection]
& (Ixy) /.x() /(M+xy) *+x(I-xy) (I-xy)
(I'(12345abcdef)'(abcdefUVWXYZ))
[Subtract set y from set x]
& (Sxy) /.x() /(M+xy) (S-xy) *+x(S-xy)
(S'(12345abcdef)'(abcdefUVWXYZ))
[Identity function that outputs a set of elements]
& (Os) /.s() *,+s(O-s)
(O'(12345abcdef))
[Combine two bit strings by interleaving them]
& (Cxy) /.xy /.yx *+x*+y(C-x-y)
(C'(0000000000)'(11111111111111111111))
\end{verbatim}
}\chap{sets1.lisp}{\size\begin{verbatim}
[[[
 Show that
    H(X set-union Y) <= H(X) + H(Y) + c
 and that
    H(X set-intersection Y) <= H(X) + H(Y) + c
 and determine both c's.
 Here X and Y are INFINITE sets.
]]]

[Combine two bit strings by interleaving them]
& (Cxy) /.xy /.yx *+x*+y(C-x-y)

[[[++]]]?0'!%

^,#,'

[Package of set functions from sets0.lisp]
: (Mes) /.s0 /=e+s1 (Me-s)
: (Ds) /.s() /(M+s-s) (D-s) *+s(D-s)
: (Uxy) /.xy /(M+xy) (U-xy) *+x(U-xy)
: (Ixy) /.x() /(M+xy) *+x(I-xy) (I-xy)
: (Sxy) /.x() /(M+xy) (S-xy) *+x(S-xy)
: (Os) /.s() *~,+s(O-s) [<===cheating to get display!]
[Main Loop:
 o is set of elements output so far.
 For first set,
 t is depth limit (time), b is bits read so far.
 For second set,
 u is depth limit (time), c is bits read so far.
]
: (Lotbuc)
 : v     ~?~t'!%~b   [Run 1st computation again.]
 : w     ~?~u'!%~c   [Run 2nd computation again.]
 : x     (U-v-w)     [Form UNION of sets so far]
 : y     (O(Sxo))    [Output all new elements]
 [This is an infinite loop. But to make debugging easier,
  halt if both computations have halted.]
 / /=0.+v /=0.+w 100 x [If halts, value is output so far]
 [Bump everything before branching to head of loop]
 (L x                [Value of y is discarded, x is new o]
    /="?+v *1t t     [Increment time for 1st computation]
    /="!+v  ^b*@() b [Increment tape for 1st computation]
    /="?+w *1u u     [Increment time for 2nd computation]
    /="!+w  ^c*@() c [Increment tape for 2nd computation]
 )

(L()()()()())        [Initially all 5 induction variables
                      are empty.]

,
(C                   [Combine their bits in order needed]
                     [Wrong order if programs use @ or %]
,#,'*,a*,b*,c0       [Program to enumerate 1st FINITE set]
,#,'*,b*,c*,d0       [Program to enumerate 2nd FINITE set]
)
\end{verbatim}
}\chap{sets2.lisp}{\size\begin{verbatim}
[[[
 Show that
    H(X set-union Y) <= H(X) + H(Y) + c
 and that
    H(X set-intersection Y) <= H(X) + H(Y) + c
 and determine both c's.
 Here X and Y are INFINITE sets.
]]]

[Combine two bit strings by interleaving them]
& (Cxy) /.xy /.yx *+x*+y(C-x-y)

[[[++]]]?0'!%

^,#,'

[Package of set functions from sets0.lisp]
: (Mes) /.s0 /=e+s1 (Me-s)
: (Ds) /.s() /(M+s-s) (D-s) *+s(D-s)
: (Uxy) /.xy /(M+xy) (U-xy) *+x(U-xy)
: (Ixy) /.x() /(M+xy) *+x(I-xy) (I-xy)
: (Sxy) /.x() /(M+xy) (S-xy) *+x(S-xy)
: (Os) /.s() *~,+s(O-s) [<===cheating to get display!]
[Main Loop:
 o is set of elements output so far.
 For first set,
 t is depth limit (time), b is bits read so far.
 For second set,
 u is depth limit (time), c is bits read so far.
]
: (Lotbuc)
 : v     ~?~t'!%~b   [Run 1st computation again.]
 : w     ~?~u'!%~c   [Run 2nd computation again.]
 : x     (I-v-w)     [Form INTERSECTION of sets so far]
 : y     (O(Sxo))    [Output all new elements]
 [This is an infinite loop. But to make debugging easier,
  halt if both computations have halted.]
 / /=0.+v /=0.+w 100 x [If halts, value is output so far]
 [Bump everything before branching to head of loop]
 (L x                [Value of y is discarded, x is new o]
    /="?+v *1t t     [Increment time for 1st computation]
    /="!+v  ^b*@() b [Increment tape for 1st computation]
    /="?+w *1u u     [Increment time for 2nd computation]
    /="!+w  ^c*@() c [Increment tape for 2nd computation]
 )

(L()()()()())        [Initially all 5 induction variables
                      are empty.]

,
(C                   [Combine their bits in order needed]
                     [Wrong order if programs use @ or %]
,#,'*,a*,b*,c0       [Program to enumerate 1st FINITE set]
,#,'*,b*,c*,d0       [Program to enumerate 2nd FINITE set]
)
\end{verbatim}
}\chap{sets3.lisp}{\size\begin{verbatim}
[[[
 Show that
    H(X set-union Y) <= H(X) + H(Y) + c
 and that
    H(X set-intersection Y) <= H(X) + H(Y) + c
 and determine both c's.
 Here X and Y are INFINITE sets.
]]]

[Combine two bit strings by interleaving them]
& (Cxy) /.xy /.yx *+x*+y(C-x-y)

[IMPORTANT: This test case never halts, so] [<=====!!!]
[must run this with xclisp.m, not clisp.m.]

[[[++]]]?0'!%

^,#,'

[Package of set functions from sets0.lisp]
: (Mes) /.s0 /=e+s1 (Me-s)
: (Ds) /.s() /(M+s-s) (D-s) *+s(D-s)
: (Uxy) /.xy /(M+xy) (U-xy) *+x(U-xy)
: (Ixy) /.x() /(M+xy) *+x(I-xy) (I-xy)
: (Sxy) /.x() /(M+xy) (S-xy) *+x(S-xy)
: (Os) /.s() *~,+s(O-s) [<===cheating to get display!]
[Main Loop:
 o is set of elements output so far.
 For first set,
 t is depth limit (time), b is bits read so far.
 For second set,
 u is depth limit (time), c is bits read so far.
]
: (Lotbuc)
 : v     ?t'!%b      [Run 1st computation again.]
 : w     ?u'!%c      [Run 2nd computation again.]
 : x     (I-v-w)     [Form INTERSECTION of sets so far]
 : y     (O(Sxo))    [Output all new elements]
 [This is an infinite loop. But to make debugging easier,
  halt if both computations have halted.]
 / /=0.+v /=0.+w 100 x [If halts, value is output so far]
 [Bump everything before branching to head of loop]
 (L x                [Value of y is discarded, x is new o]
    /="?+v *1t t     [Increment time for 1st computation]
    /="!+v  ^b*@() b [Increment tape for 1st computation]
    /="?+w *1u u     [Increment time for 2nd computation]
    /="!+w  ^c*@() c [Increment tape for 2nd computation]
 )

(L()()()()())        [Initially all 5 induction variables
                      are empty.]

,
(C                   [Combine their bits in order needed]
                     [Wrong order if programs use @ or %]
                     [Program to enumerate 1st INFINITE set]
,#,':(Lk)(L,*1*1k)(L())
                     [Program to enumerate 2nd INFINITE set]
,#,':(Lk)(L,*1*1*1k)(L())
)
\end{verbatim}
}\chap{sets4.lisp}{\size\begin{verbatim}
[[[
 Show that
    H(X set-union Y) <= H(X) + H(Y) + c
 and that
    H(X set-intersection Y) <= H(X) + H(Y) + c
 and determine both c's.
 Here X and Y are INFINITE sets.
]]]

[Combine two bit strings by interleaving them]
& (Cxy) /.xy /.yx *+x*+y(C-x-y)

[IMPORTANT: This test case never halts, so] [<=====!!!]
[must run this with xclisp.m, not clisp.m.]

[[[++]]]?0'!%

^,#,'

[Package of set functions from sets0.lisp]
: (Mes) /.s0 /=e+s1 (Me-s)
: (Ds) /.s() /(M+s-s) (D-s) *+s(D-s)
: (Uxy) /.xy /(M+xy) (U-xy) *+x(U-xy)
: (Ixy) /.x() /(M+xy) *+x(I-xy) (I-xy)
: (Sxy) /.x() /(M+xy) (S-xy) *+x(S-xy)
: (Os) /.s() *~,+s(O-s) [<===cheating to get display!]
[Main Loop:
 o is set of elements output so far.
 For first set,
 t is depth limit (time), b is bits read so far.
 For second set,
 u is depth limit (time), c is bits read so far.
]
: (Lotbuc)
 : v     ?t'!%b      [Run 1st computation again.]
 : w     ?u'!%c      [Run 2nd computation again.]
 : x     (U-v-w)     [Form UNION of sets so far]
 : y     (O(Sxo))    [Output all new elements]
 [This is an infinite loop. But to make debugging easier,
  halt if both computations have halted.]
 / /=0.+v /=0.+w 100 x [If halts, value is output so far]
 [Bump everything before branching to head of loop]
 (L x                [Value of y is discarded, x is new o]
    /="?+v *1t t     [Increment time for 1st computation]
    /="!+v  ^b*@() b [Increment tape for 1st computation]
    /="?+w *1u u     [Increment time for 2nd computation]
    /="!+w  ^c*@() c [Increment tape for 2nd computation]
 )

(L()()()()())        [Initially all 5 induction variables
                      are empty.]

,
(C                   [Combine their bits in order needed]
                     [Wrong order if programs use @ or %]
                     [Program to enumerate 1st INFINITE set]
,#,':(Lk)(L,*1*1k)(L())
                     [Program to enumerate 2nd INFINITE set]
,#,':(Lk)(L,*2*2*2k)(L())
)
\end{verbatim}
}\chap{godel.lisp}{\size\begin{verbatim}
[[[
 Show that a formal system of complexity N
 can't prove that a specific object has
 complexity > N + c, and also determine c.
 Formal system is a never halting lisp expression
 that output pairs (lisp object, lower bound
 on its complexity).  E.g., (x(1111)) means
 that x has complexity H(x) greater than 4.
]]]

[ (<xy) tells if x is less than y ]
& (<xy) /.x /.y01
            /.y0(<-x-y)
(<'(11)'(11))
(<'(11)'(111))
(<'(111)'(11))

[ Examine pairs in p to see if 2nd is greater than n ]
[ returns 0 to indicate not found, or pair if found ]
& (Epn) /.p 0 /(<n+-+p) +p (E-pn)
(E'((x(11))(y(111)))'())
(E'((x(11))(y(111)))'(1))
(E'((x(11))(y(111)))'(11))
(E'((x(11))(y(111)))'(111))
(E'((x(11))(y(111)))'(1111))

++?0'!%

^,#,'
[ (<xy) tells if x is less than y ]
: (<xy) /.x /.y01
            /.y0(<-x-y)
[ Over-write real definition for test ]
: (<xy) 1

[ Examine pairs in p to see if 2nd is greater than n ]
[ returns 0 to indicate not found, or pair if found ]
: (Epn) /.p 0 /(<n+-+p) +p (E-pn)

[Parameter in proof]
: k ~'(11111)
: k ~ ^kk
: k ~ ^kk
: k ~ ^kk
: k ~ ^kk

[Main Loop - t is depth limit (time), b is bits read so far]
: (Ltb)
 : v ~?~t'!%~b [run universal computer again]
 : s (E-v^kb) [look for pair with 2nd > 16k + # of bits read]
 /s +s       [Found it!  Output 1st and halt]
 /="!+v  (Lt^b*@()) [Read another bit from program tape]
 /="?+v  (L*1tb)    [Increase depth/time limit]
 "?     [Surprise, formal system halts, so we do too]

(L()())    [Initially, 0 depth limit and no bits read]

[
,#,','((xy)(11))
]
,#,','(x())
\end{verbatim}
}\chap{godel2.lisp}{\size\begin{verbatim}
[[[
 Show that a formal system of complexity N
 can't prove that a specific object has
 complexity > N + c, and also determine c.
 Formal system is a never halting lisp expression
 that output pairs (lisp object, lower bound
 on its complexity).  E.g., (x(1111)) means
 that x has complexity H(x) greater than 4.
]]]

++?0'!%

^,#,'
[ (<xy) tells if x is less than y ]
: (<xy) /.x /.y01
            /.y0(<-x-y)

[ Examine pairs in p to see if 2nd is greater than n ]
[ returns 0 to indicate not found, or pair if found ]
: (Epn) /.p 0 /(<n+-+p) +p (E-pn)

[Parameter in proof]
: k ~'(11111)
: k ~ ^kk
: k ~ ^kk
: k ~ ^kk
: k ~ ^kk

[Main Loop - t is depth limit (time), b is bits read so far]
: (Ltb)
 : v ~?~t'!%~b [run universal computer again]
 : s (E-v^kb) [look for pair with 2nd > 16k + # of bits read]
 /s +s       [Found it!  Output 1st and halt]
 /="!+v  (Lt^b*@()) [Read another bit from program tape]
 /="?+v  (L*1tb)    [Increase depth/time limit]
 "?     [Surprise, formal system halts, so we do too]

(L()())    [Initially, 0 depth limit and no bits read]

[
,#,','((xy)(11))
]
,#,','(x())
\end{verbatim}
}\chap{godel3.lisp}{\size\begin{verbatim}
[[[
 Show that a formal system of complexity N
 can't determine more than N + c bits of Omega,
 and also determine c.
 Formal system is a never halting lisp expression
 that outputs lists of the form (10X0XXXX10).
 This stands for the fractional part of Omega,
 and means that these 0,1 bits of Omega are known.
 X stands for an unknown bit.
]]]

[Count number of bits in an omega that are determined.]
& (Cw) /.w() ^ /=0+w'(1) /=1+w'(1) ()
               (C-w)
(C'(XXX))
(C'(1XX))
(C'(1X0))
(C'(110))

[Merge bits of data into unknown bits of an omega.]
& (Mw) /.w() * /=0+w0 /=1+w1 @
               (M-w)
[Test it.]
++?0 ':(Mw)/.w()*/=0+w0/=1+w1@(M-w) (M'(00X00X00X)) '(111)
++?0 ':(Mw)/.w()*/=0+w0/=1+w1@(M-w) (M'(11X11X111)) '(00)

[(<xy) tells if x is less than y.]
& (<xy) /.x /.y01
            /.y0(<-x-y)
(<'(11)'(11))
(<'(11)'(111))
(<'(111)'(11))

[
 Examine omegas in list w to see if in any one of them
 the number of bits that are determined is greater than n.
 Returns 0 to indicate not found, or what it found.
]
& (Ewn) /.w 0 /(<n(C+w)) +w (E-wn)
(E'((00)(000))'())
(E'((00)(000))'(1))
(E'((00)(000))'(11))
(E'((00)(000))'(111))
(E'((00)(000))'(1111))

++?0'!%

^,#,'
[Count number of bits in an omega that are determined.]
: (Cw) /.w() ^ /=0+w'(1) /=1+w'(1) ()
               (C-w)

[Merge bits of data into unknown bits of an omega.]
: (Mw) /.w() * /=0+w0 /=1+w1 @
               (M-w)

[(<xy) tells if x is less than y.]
: (<xy) /.x /.y01
            /.y0(<-x-y)
[Over-write real definition for test.]
: (<xy) 1

[
 Examine omegas in list w to see if in any one of them
 the number of bits that are determined is greater than n.
 Returns 0 to indicate not found, or what it found.
]
: (Ewn) /.w 0 /(<n(C+w)) +w (E-wn)

[Parameter in proof]
: k ~'(11111)
: k ~ ^kk
: k ~ ^kk
: k ~ ^kk
: k ~ ^kk

[Main Loop: t is depth limit (time), b is bits read so far.]
: (Ltb)
 : v     ~?~t'!%~b  [Run universal computer again.]
 : s     (E-v^kb)   [Look for an omega with >
                     (16k + # of bits read) bits determined.]
 /s      (Ms)       [Found it!  Merge in undetermined bits,
                     output result, and halt.]
 /="!+v  (Lt^b*@()) [Read another bit from program tape.]
 /="?+v  (L*1tb)    [Increase depth/time limit.]
         "?         [Surprise, formal system halts,
                     so we do too.]

(L()())             [Initially, 0 depth limit
                     and no bits read.]

^,#,'
,'(1X0) [Toy formal system with only one theorem.]

,'
(0) [Missing bit of omega that is needed.]
\end{verbatim}

\part{The Software}

}\chap{slisp.m}{\size\begin{verbatim}
(***** SLISP.M *****)

<<frontend.m

(* string lisp interpreter *)

chars2bits = (* convert chars to binary *)
(FromCharacterCode@ # -> StringJoin[
 Rest@ IntegerDigits[128 + #, 2] /.
 {0 -> "0", 1 -> "1"} ]
) & /@ Range[0,127]

bits2chars = chars2bits /. (l_->r_)->(r->l)

getbit[] :=
Block[ {x},
 trouble = False; (* Mma bug bypass *)
 If[ at@ tape, (trouble = True; Throw@ "!")];
 x = hd@ tape;
 tape = tl@ tape;
 If[ x === "0", "0", "1" ]
]

getchar[] := (* nonprintables -> "?" *)
 FromCharacterCode[
 If[ 31 < # < 127, #, 63 ]& /@
 ToCharacterCode[ StringReplace[
  getbit[]<>getbit[]<>getbit[]<>getbit[]<>
  getbit[]<>getbit[]<>getbit[],
  bits2chars
]]]

getexp[rparenokay_:False] :=
Block[ { c = getchar[], d, l = "(" },
 Switch[
 c,
 ")", Return@ If[rparenokay,")","()"],
 "(",
(While[ ")" =!= (d = getexp[True]),
 l = l <> d
 ];
 If[ trouble, Throw@ "!" ]; (* Mma bug bypass *)
 Return@ ( l <> ")" )
),
 _, c
 ]
]

at[x_] := StringLength@ x == 1 || x === "()"

hd[x_] :=
(If[ at@ x, Return@ x ];
 Block[ {p = 0},
 Do[
 p += Switch[ StringTake[x,{i}], "(", +1, ")", -1, _, 0 ];
 If[ p == 0, Return@ StringTake[x,{2,i}] ],
 {i, 2, StringLength@ x}
 ]
 ]
)

tl[x_] :=
(If[ at@ x, Return@ x ];
 Block[ {p = 0},
 Do[
 p += Switch[ StringTake[x,{i}], "(", +1, ")", -1, _, 0 ];
 If[ p == 0, Return[ "("<>StringDrop[x,i] ] ],
 {i, 2, StringLength@ x}
 ]
 ]
)

jn[x_,y_] :=
 If[ StringLength@ y == 1, x, "("<>x<>StringDrop[y,1] ]

eval[e_,,d_] := eval[e,"()",d]

eval[e2_,a_,d2_] :=

Block[ {e = e2, d = d2, f, args, x, y, z},
 If[
 at@ e,
 Return@
 Which[
 e === hd@ a, hd@tl@ a,
 at@ a, e,
 True, eval[ e, tl@tl@ a, ]
 ]
 ];
 f = eval[ hd@ e, a, d ];
 e = tl@ e;
 Switch[
 f,
 "'", Return@ hd@ e,
 "/", Return@
 If[
 eval[hd@ e,a,d] =!= "0",
 eval[hd@tl@ e,a,d],
 eval[hd@tl@tl@ e,a,d]
 ]
 ];
 args = evlst[e,a,d];
 x = hd@ args;
 y = hd@tl@ args;
 z = hd@tl@tl@ args;
 Switch[
 f,
 "@", Return@ getbit[],
 "%", Return@ getexp[],
 "#", Return@ ( "(" <> StringReplace[x,chars2bits] <> ")" ),
 "+", Return@ hd@ x,
 "-", Return@ tl@ x,
 "*", Return@ jn[x,y],
 "^", Return@ ( StringDrop[ If[at@x,"()",x], -1 ] <>
                StringDrop[ If[at@y,"()",y],  1 ]
              ),
 ".", Return@ If[ at@ x, "1", "0" ],
 "=", Return@ If[ x === y, "1", "0" ],
 ",", Return@ ( out = jn[x,out];
                If[ display, print[ "display", output@ x ] ];
                x
              ),
 "~", Return@ ( print[ "display", output@ x ];
                x
              )
 ];
 If[ d == 0, Throw@ "?" ];
 d--;
 Switch[
 f,
 "!", Return@ eval[x,,d],
 "?", Return@
 Block[ {out = "()", tape = z, display = False},
 jn[ If[ size@x < d,
 Catch[ "("<>eval[y,,size@x]<>")" ],
 Catch[ "("<>eval[y,,d]<>")" ] //
 If[ # === "?", Throw@ #, # ] &
 ], out ] ]
 ];
 f = tl@ f;
 eval[ hd@tl@ f, bind[hd@ f,args,a], d ]
]

size["()"] := 0
size[x_?at] := Infinity
size[x_] := 1 + size@ tl@ x

evlst[e_?at,a_,d_] := e
evlst[e_,a_,d_] := jn[ eval[hd@ e,a,d], evlst[tl@ e,a,d] ]

bind[vars_?at,args_,a_] := a
bind[vars_,args_,a_] :=
 jn[hd@ vars, jn[hd@ args, bind[tl@ vars,tl@ args,a]]]

eval[e_] :=
(
 out = tape = "()";
 display = True;
 print[ "expression", output@ e ];
 Catch[ eval[ output@ wrap@ e,,Infinity ] ]
)

run[fn_] := run[fn, "slisp.m", ".srun"]
\end{verbatim}
}\chap{lisp.m}{\size\begin{verbatim}
(***** LISP.M *****)

<<frontend.m

(* fast lisp interpreter *)

chars2bits = (* convert chars to binary *)
(FromCharacterCode@ # -> StringJoin[
 Rest@ IntegerDigits[128 + #, 2] /.
 {0 -> "0", 1 -> "1"} ]
) & /@ Range[0,127]

bits2chars = chars2bits /. (l_->r_)->(r->l)

getbit[] :=
Block[ {x},
 trouble = False; (* Mma bug bypass *)
 If[ at@ tape, (trouble = True; Throw@ "!")];
 x = hd@ tape;
 tape = tl@ tape;
 If[ x === "0", "0", "1" ]
]

getchar[] := (* nonprintables -> "?" *)
 FromCharacterCode[
 If[ 31 < # < 127, #, 63 ]& /@
 ToCharacterCode[ StringReplace[
  getbit[]<>getbit[]<>getbit[]<>getbit[]<>
  getbit[]<>getbit[]<>getbit[],
  bits2chars
]]]

getexp[rparenokay_:False] :=
Block[ { c = getchar[], d, l = {} },
 Switch[
 c,
 ")", Return@ If[rparenokay,")",{}],
 "(",
(While[ ")" =!= (d = getexp[True]),
 AppendTo[l,d]
 ];
 If[ trouble, Throw@ "!"]; (* Mma bug bypass *)
 Return@ l
),
 _, c
 ]
]

identitymap =
 ( FromCharacterCode /@ Range[0,127] ) ~Join~ {{},}

pos[c_String] :=
 ( If[ # <= 128, #, Abort[] ] )& @
 ( 1 + First@ ToCharacterCode@ c )
pos[{}] :=
 129
pos[_] :=
 130

at[x_] :=
 MatchQ[ x, {}|_String ]
hd[x_] :=
 If[ at@ x, x, First@ x ]
tl[x_] :=
 If[ at@ x, x, Rest@ x ]
jn[x_,y_] :=
 If[ MatchQ[y,_String], x, Prepend[y,x] ]

eval[e_,,d_] := eval[e,identitymap,d]

eval[e2_,a_,d2_] :=

Block[ {e = e2, d =d2, f, args, x, y, z},
 If[ at@ e, Return@ a[[ pos@ e ]] ];
 f = eval[hd@ e,a,d];
 e = tl@ e;
 Switch[
 f,
 "'", Return@ hd@ e,
 "/", Return@
 If[
 eval[hd@ e,a,d] =!= "0",
 eval[hd@tl@ e,a,d],
 eval[hd@tl@tl@ e,a,d]
 ]
 ];
 args = eval[#,a,d]& /@ e;
 x = hd@ args;
 y = hd@tl@ args;
 z = hd@tl@tl@ args;
 Switch[
 f,
 "@", Return@ getbit[],
 "%", Return@ getexp[],
 "#", Return@ Characters@ StringReplace[output@x,chars2bits],
 "+", Return@ hd@ x,
 "-", Return@ tl@ x,
 "*", Return@ jn[x,y],
 "^", Return@ Join[ If[at@x,{},x], If[at@y,{},y] ],
 ".", Return@ If[ at@ x, "1", "0" ],
 "=", Return@ If[ x === y, "1", "0" ],
 ",", Return@ (out = jn[x,out];
               If[ display, print[ "display", output@ x ] ];
               x),
 "~", Return@ (print[ "display", output@ x ];
               x)
 ];
 If[ d == 0, Throw@ "?" ];
 d--;
 Switch[
 f,
 "!", Return@ eval[x,,d],
 "?", Return@
 Block[ {out = {}, tape = z, display = False},
 jn[ If[ size@x < d,
 Catch@ {eval[y,,size@x]},
 Catch@ {eval[y,,d]} //
 If[ # === "?", Throw@ #, # ] &
 ], out ] ]
 ];
 f = tl@ f;
 eval[ hd@tl@ f, bind[hd@ f,args,a], d ]
]

size[x_?at] := If[ x === {}, 0, Infinity ]
size[x_] :=  Length@x

bind[vars_?at,args_,a_] :=
 a

bind[vars_,args_,a_] :=
ReplacePart[
 bind[ tl@ vars, tl@ args, a ],
 hd@ args,
 pos@ hd@ vars
]

eval[e_] :=
(
 out = tape = {};
 display = True;
 print[ "expression", output@ e ];
 Catch[ eval[ wrap@ e,,Infinity ] ]
)

run[fn_] := run[fn, "lisp.m", ".run"]
\end{verbatim}
}\chap{lispm.m}{\size\begin{verbatim}
(***** LISPM.M *****)

<<frontend.m

(* lisp machine interpreter *)

p = << lisp.xrm

labels = Cases[p, {l_,__} -> l]

If[
 Length@ Union@ labels != Length@ p,
 Print@ "Duplicate labels!!!"
]

registers = Cases[p, {_,_,r__} -> r] // Flatten // Union
registers = Cases[registers, r_Symbol -> r]
registers = Complement[registers,labels]

Evaluate[ next /@ labels ] = RotateLeft@ labels
Evaluate[ #[]& /@ registers ] = {}& /@ registers
Evaluate[ #[]& /@ labels ] =
 Cases[p, {l_,op_,x___} -> op[next[l],x]]

first[x_] := If[ x === {}, "\0", x[[1]] ]

out[n_,r_] :=
(
 print[ "display", StringJoin@@ Flatten@ r[] ];
 n
)

dump[n_] :=
(
 print[ ToString@ #, StringJoin@@ Flatten@ #[] ] & /@
 registers;
 n
)

eqi[n_,r_,i_,l_] := If[ first[r[]] === i, l, n ]
neqi[n_,r_,i_,l_] := If[ first[r[]] =!= i, l, n ]
eq[n_,r_,s_,l_] := If[ first[r[]] === first[s[]], l, n ]
neq[n_,r_,s_,l_] := If[ first[r[]] =!= first[s[]], l, n ]

lefti[n_,r_,i_] :=
If[
 i === "\0", error[],
 r[] = {i, r[]};
 n
]

left[n_,r_,s_] :=
If[
 s[] === {}, error[],
 r[] = {s[][[1]], r[]};
 s[] = s[][[2]];
 n
]

right[n_,r_] :=
If[
 r[] === {}, error[],
 r[] = r[][[2]];
 n
]

seti[n_,r_,"\0"] := (r[] = {}; n)
seti[n_,r_,i_] := (r[] = {i, {}}; n)
set[n_,r_,s_] := (r[] = s[]; n)

goto[n_,l_] := l
halt[n_] := halt
error[] := (Print@ "ERROR!!!"; Abort[])

ravel[c_,r___] := {c, ravel[r]}
ravel[] := {}

jump[n_,r_,l_] :=
(
 r[] = ravel@@ Characters[ "("<>ToString[n]<>")" ];
 l
)

goback[n_,r_] :=
ToExpression[
 StringJoin@@ Drop[ Drop[ Flatten@ r[], 1], -1]
]

eval[e_] :=
(
 print[ "expression", output@ e ];
 reg$expression[] = ravel@@ Characters@
    output@ wrap@ e;
 reg$all$ascii$characters[] = ravel@@ Characters@
    FromCharacterCode@ Range[127,0,-1];
 reg$printable$ascii$characters[] =  ravel@@ Characters@
    FromCharacterCode[
       If[ 31 < # < 127, #, 63 ]& /@ Range[127,0,-1]
    ];
 reg$ascii$bits[] = ravel@@ Flatten[
    (
    Rest@ IntegerDigits[ 128 + #, 2 ] /. {0 -> "0", 1 -> "1"}
    )&
    /@
    Range[127,0,-1]
 ];
 loc = lab$l1;
 While[ loc =!= halt, clock++; loc = loc[] ];
 StringJoin@@ Flatten@ reg$value[]
)

run[fn_] := run[fn, "lispm.m", ".mrun"]
\end{verbatim}
}\chap{clisp.m}{\size\begin{verbatim}
(* CLISP.M *)

<<frontend.m

(* driver for C lisp interpreter *)

eval[e_] :=
(
 print[ "expression", output@ e ];
 t1 = "tmp1"<>ToString@ Random[Integer,10^10];
 t2 = "tmp2"<>ToString@ Random[Integer,10^10];
 tmp1 = OpenWrite@ t1;
 WriteString[tmp1, output@ wrap@ e,"\n"];
 Close@ tmp1;
 Run["lisp","<",t1,">",t2];
 tmp2 = ReadList[t2,Record];
 Run["rm",t1];
 Run["rm",t2];
 print["display",#]& /@ Drop[tmp2,-1];
 tmp2[[-1]]
)

run[fn_] := run[fn, "clisp.m", ".crun" ]
\end{verbatim}
}\chap{xclisp.m}{\size\begin{verbatim}
(* XCLISP.M *)

<<frontend.m

(* driver for C lisp interpreter *)
(* allows C to do output directly *)

eval[e_] :=
(
 print[ "expression", output@ e ];
 t1 = "tmp1"<>ToString@ Random[Integer,10^10];
 tmp1 = OpenWrite@ t1;
 WriteString[tmp1, output@ wrap@ e,"\n"];
 Close@ tmp1;
 Run["lisp","<",t1];
 Run["rm",t1];
 " "
)

run[fn_] := run[fn, "xclisp.m", ".xcrun" ]
\end{verbatim}
}\chap{clispm.m}{\size\begin{verbatim}
(* CLISPM.M *)

<<frontend.m

(* driver for C lisp machine *)

eval[e_] :=
(
 print[ "expression", output@ e ];
 t1 = "tmp1"<>ToString@ Random[Integer,10^10];
 t2 = "tmp2"<>ToString@ Random[Integer,10^10];
 tmp1 = OpenWrite@ t1;
 WriteString[tmp1, StringReverse@ output@ wrap@ e,"\n"];
 Close@ tmp1;
 Run["lispm","<",t1,">",t2];
 tmp2 = ReadList[t2,Record];
 Run["rm",t1];
 Run["rm",t2];
 clock = ToExpression@ tmp2[[-1]];
 tmp2 = StringReverse /@ Drop[tmp2,-1];
 print["display",#]& /@ Drop[tmp2,-1];
 tmp2[[-1]]
)

run[fn_] := run[fn, "clispm.m", ".cmrun"]
\end{verbatim}
}\chap{frontend.m}{\size\begin{verbatim}
(***** FRONTEND.M *****)

chr2[]:=
Block[ {c},
While[
 StringLength@ line == 0,
 line = Read[i,Record];
 If[ line == EndOfFile, Abort[] ];
 Print@ line;
 WriteString[o,line,"\n"];
 (* keep only non-blank printable ASCII codes *)
 line = FromCharacterCode@
 Cases[ ToCharacterCode@ line, n_Integer /; 32 < n < 127 ]
];
c = StringTake[line,1];
line = StringDrop[line,1];
c
]

chr[] :=
Block[ {c},
While[ True,
 c = chr2[];
 If[ c =!= "[", Return@ c ];
 While[ chr[] =!= "]" ]
]
]

get[sexp_:False,rparenokay_:False] :=

Block[ {c = chr[], d, l ={}, name, def, body, varlist},
 Switch[
 c,
 ")", Return@ If[rparenokay,")",{}],
 "(",
 While[ ")" =!= (d = get[sexp,True]),
 AppendTo[l,d]
 ];
 Return@ l
 ];
 If[ sexp, Return@ c ];
 Switch[
 c,
 "\"", get[True],
 ":",
 {name,def,body} = {get[],get[],get[]};
 If[
 !MatchQ[name,{}|_String],
 varlist = Rest@ name;
 name = First@ name;
 def = {"'",{"&",varlist,def}}
 ];
 {{"'",{"&",{name},body}},def},
 "@"|"%", {c},
 "+"|"-"|"."|"'"|","|"!"|"#"|"~", {c,get[]},
 "*"|"="|"&"|"^", {c,get[],get[]},
 "/"|":"|"?", {c,get[],get[],get[]},
 _, c
 ]
]

(* output S-exp *)
output[x_String] := x
output[{x___}] := StringJoin["(", output /@ {x}, ")"]

blanks = StringJoin@ Table[" ",{12}]

print[x_,y_] := (print2[x,StringTake[y,50]];
 print["",StringDrop[y,50]]) /; StringLength[y] > 50
print[x_,y_] := print2[x,y]
print2[x_,y_] := print3[StringTake[x<>blanks,12]<>y]
print3[x_] := (Print[x]; WriteString[o,x,"\n"])

wrap[e_] :=
If[ names === {}, e, {{"'",{"&",names,e}}} ~Join~ defs ]

let[n_,d_] :=
(
 print[ output@ n<> ":", output@ d ];
 names = {n} ~Join~ names;
 defs = {{"'",d}} ~Join~ defs;
)

run[fn_,whoami_,outputsuffix_] :=
(
 line = "";
 names = defs = {};
 t0 = SessionTime[];
 o = OpenWrite[fn<>outputsuffix];
 i = OpenRead[fn<>".lisp"];
 print3["Start of "<>whoami<>" run of "<>fn<>".lisp"];
 print3@ "";
 CheckAbort[
 While[True,
(print3@ "";
 clock = 0;
 Replace[#,{
 {"&",{func_,vars___},def_} :> let[func,{"&",{vars},def}],
 {"&",var_,def_} :> let[var,eval@ def],
 _ :> print[ "value", output@ eval@ # ]
 }];
 If[clock != 0, print["cycles",ToString@clock]]
)& @ get[];
 print3@ ""
 ],
 ];
 print3@ StringForm[
 "Elapsed time `` seconds",
 Round[SessionTime[]-t0]
 ];
 Close@ i;
 Close@ o
)

runall := run /@ {"univ",
                 "omega","omega2","omega3","omega4",
                 "godel","godel2","godel3",
                 "sets0","sets1","sets2"}
                 (* sets3 & 4 don't halt *)

$RecursionLimit = $IterationLimit = Infinity
SetOptions[$Output,PageWidth->63];
\end{verbatim}
}\chap{xpnd.m}{\size\begin{verbatim}
(***** XPND.M *****)

Off[ General::spell, General::spell1 ]

run[fn_String] := Module[ {p, o},

(* program p is list of instructions of form: l, op[r,s], *)
p = Get[fn<>".rm"];

SetOptions[$Output,PageWidth->62];
Format[LineBreak[_]] = "";
Format[Continuation[_]] = " ";
Print@ "(**** before ****)";
Print@ Short[InputForm@p,10];

p = p //. {
set[x_,x_] ->
 {},
split[h_,t_,s_] ->
 {set[source,s], jump[linkreg3,split$routine],
 set[h,target], set[t,target2]},
hd[t_,s_] ->
 split[t,target2,s],
tl[t_,s_] ->
 split[target,t,s],
empty[r_] ->
 {set[r,")"], left[r,"("]},
atom[r_,l_] ->
 {neq[r,"(",l], set[work,r], right[work], eq[work,")",l]},
jn[i_,x_,y_] ->
 {set[source,x], set[source2,y], jump[linkreg3,jn$routine],
 set[i,target]},
push[x_] ->
 {set[source,x], jump[linkreg2,push$routine]},
pop[x_] ->
 {jump[linkreg2,pop$routine], set[x,target]},
popl[x_,y_] ->
 split[x,y,y]
};

p = Flatten@ p;

p = p /. op_[l___, x_String, r___]
 :> ( ToExpression[ ToString@ op<> "i" ] )[l,x,r];

p = p //. {l___, x_Symbol, y_, r___}
 -> {l, label[x,y], r};

labels =
 ( ToExpression[ "l"<> ToString@ # ] )& /@ Range@ Length@ p;

p = MapThread[ Replace[#1,
 {label[x__] -> label[x], x_ -> label[#2,x]} ]&,
 {p,labels} ];

p = p /. label[x_,op_[y___]] -> {x,op,y};

r[x_] := ToExpression["reg$"<> ToString@ x]; (* register *)
l[x_] := ToExpression["lab$"<> ToString@ x]; (* label *)
i[x_] := x; (* immediate field *)

t[x_] := x /. {
 {a_,op:halt|dump} :> {l@ a, op},
 {a_,op:goto,b_} :> {l@ a, op, l@ b},
 {a_,op:jump,b_,c_} :> {l@ a, op, r@ b, l@ c},
 {a_,op:goback|right|out,b_} :> {l@ a, op, r@ b},
 {a_,op:eq|neq,b_,c_,d_} :> {l@ a, op, r@ b, r@ c, l@ d},
 {a_,op:eqi|neqi,b_,c_,d_} :> {l@ a, op, r@ b, i@ c, l@ d},
 {a_,op:left|set,b_,c_} :> {l@ a, op, r@ b, r@ c},
 {a_,op:lefti|seti,b_,c_} :> {l@ a, op, r@ b, i@ c} };

p = t /@ p;

Print@ "(**** after ****)";
Print@ Short[InputForm@p,10];

o = OpenWrite[fn<>".xrm",PageWidth->62];
Write[o,p];
Close@ o

]

runall := run /@ {"lisp"}
\end{verbatim}
}\chap{rm2c.m}{\size\begin{verbatim}
(* RM2C.M *)

p = <<lisp.xrm
p = (ToString /@ #)& /@ p
p = p /. {"'" -> "\\'", "\0" -> "\\0"}
labels = #[[1]]& /@ p
Evaluate[ next /@ labels ] = RotateLeft@ labels
registers =
 Select[ Union@ Flatten@ p, StringMatchQ[#,"reg$*"]& ]

o = OpenWrite@ "lispm.c"
put[x_] := WriteString[o,StringReplace[x,map],"\n"]

map = {}

put@ "/* LISP interpreter running on register machine */"
put@ "#include <stdio.h>"
put@ "#define size 100000"
put@ ""
put@ "main() /* lisp main program */"
put@ "{"
put@ "static char *label[] = {"
(
 map = {"R" -> #};
 put@ "\"(R)\","
)& /@ labels
put@ "\"\"}; /* end of label table */"
put@ ""
put@ "char c, *i, *j, *k;"
put@ "long n;"
put@ "double cycles = 0.0;"
put@ ""
(
 map = "R" -> #;
 put@ "char $R[size] = \"\", *R = $R;"
)& /@ registers
put@ ""
put@ "while ((c = getchar()) != '\\n') *++reg$expression = c;"
put@ ""
put@ "for (n = 0; n < 128; ++n) {"
put@ " *++reg$all$ascii$characters = n;"
put@ " c = (31 < n && n < 127 ? n : '?');"
put@ " *++reg$printable$ascii$characters = c;"
put@ " *++reg$ascii$bits = (n &  1 ? '1' : '0');"
put@ " *++reg$ascii$bits = (n &  2 ? '1' : '0');"
put@ " *++reg$ascii$bits = (n &  4 ? '1' : '0');"
put@ " *++reg$ascii$bits = (n &  8 ? '1' : '0');"
put@ " *++reg$ascii$bits = (n & 16 ? '1' : '0');"
put@ " *++reg$ascii$bits = (n & 32 ? '1' : '0');"
put@ " *++reg$ascii$bits = (n & 64 ? '1' : '0');"
put@ "}"
put@ ""

Cases[p,
 {l_,op_,a_:"",b_:"",c_:""} :>
 (map =
 {
 "L" -> l, "O" -> op, "A" -> a, "B" -> b, "C" -> c,
 "N" -> StringReverse@ next@ l
 };
 put@ ("/* L: O A,B,C */");
 put@ "L: cycles += 1.0;";
 put@ Switch[
 ToExpression@op,
 dump, "/* not supported */",
 halt, "goto termination_routine;",
 goto, "goto A;",
 goback, "k = A;\ngoto goback_routine;",
 eqi, "if (*A == 'B') goto C;",
 neqi, "if (*A != 'B') goto C;",
 eq, "if (*A == *B) goto C;",
 neq, "if (*A != *B) goto C;",
 right, "if (A != $A) --A;",
 lefti,
 "if (A == ($A+size)) goto storage_full;"
 ~StringJoin~ "\n*++A = 'B';",
 left,
 "if (A == ($A+size)) goto storage_full;"
 ~StringJoin~ "\n*++A = *B;\nif (B != $B) --B;",
 seti,
 If[ b === "\\0",
 "A = $A;",
 "*(A = ($A+1)) = 'B';"
 ],
 set,
 "A = $A;\ni = $B;\nwhile (i < B) *++A = *++i;",
 out,
 "i = $A;\nwhile (i < A) putchar(*++i);\nputchar('\\n');",
 jump,
 "A = $A;\ni = \")N(\";\nwhile ((*++A = *i++) != '(');"
 ~StringJoin~ "\ngoto B;"
 ]
 )
]

put@ ""
put@ ("goto termination_routine; " ~StringJoin~
 "/* in case fell through without halting */")
put@ ""
put@ "goback_routine: n = 0;\n"
put@ "bump_label: i = k;\nj = label[n++];"
put@ "while (*j != '\\0') if (*i-- != *j++) goto bump_label;"
put@ ""
put@ "switch (n) {"
MapThread[
 (
 map = {"L" -> #1, "I" -> #2};
 put@ "case I: goto L;"
 )&,
 {labels,ToString /@ Range[1,Length@labels]}
]

put@ "default:"
put@ "printf(\"!retsasid kcabog\");\ngoto finish;"
put@ "} /* end of switch */"
put@ ""
put@ "storage_full:"
put@ "printf(\"!lluf egarots\");"
put@ "goto finish;"
put@ ""
put@ "termination_routine:"
put@ "i = $reg$value;"
put@ "while (i < reg$value) putchar(*++i);"
put@ "finish:"
put@ "printf(\"\\n%.0f\\n\",cycles);"
put@ ""
put@ "} /* end of lisp machine! */"

Close@ o

(* compile resulting C program *)
Print@ "!cc -olispm lispm.c"
Print@ "(\"!cc -O -olispm lispm.c\" with optimization)"
!cc -olispm lispm.c
\end{verbatim}
}\chap{eq.m}{\size\begin{verbatim}
(***** EQ.M *****)

fulloutput = If[ fulloutput, True, False, False ]
fn = InputString["fn of fn.xrm file = "]
t0 = SessionTime[]
p = Get[fn<>".xrm"] (* read in program *)
o = OpenWrite[fn<>".eq",PageWidth->62]
Format[LineBreak[_]] = ""
Format[Continuation[_]] = " "
print[x_] := (Print@ x; Write[o,OutputForm@ x])

print@
 "********** program"
print@
 Short[InputForm@ p,10]

(* get set of labels of all instructions in program *)

labels = #[[1]]& /@ p

If[
 Length@ Union@ labels != Length@ p,
 print@
 "Duplicate labels!"
]

(* get set of all registers in program *)

registers = Union@ Flatten@ (Drop[#,2]& /@ p)
registers = Cases[registers,_Symbol]
registers = Complement[registers,labels]

eqs = {}
put[x_] := (Write[o,x]; eqs = {eqs,x};)
Write[o,OutputForm@
 "********** <='s & =='s as they are generated"
 ]

{
 (* generate equations for base q *)
 totalinput == Plus@@ (input[#]& /@ registers),
 numberofinstructions == Length@ p,
 longestlabel == (* with ( ) around label for jump's *)
 Max@ (StringLength["("<>ToString[#]<>")"]& /@ labels),
 q == 256^
 (totalinput+ time+ numberofinstructions+ longestlabel+ 1),
 qminus1 + 1 == q,
 1 + q i == i + q^time,
 (* label equations *)
 (# <= i)& /@ labels,
 i == Plus@@ labels,
 (* equations for starting & halting *)
 1 <= p[[1,1]],
 q^time == q Plus@@ Cases[p,{l_,halt}->l]
} // put

(* generate flow equations *)

Evaluate[ next /@ labels ] = RotateLeft@ labels

{
 Cases[ p, {l_,goto,l2_} -> q l <= l2 ],
 Cases[ p, {l_,jump,a_,l2_} -> q l <= l2 ],
 Cases[ p, {l_,goback,a_} ->
 (
 { goback <= x,
 goback <= qminus1 l,
 x <= goback + qminus1 (i-l)
 } /.
 goback -> goback[l] /.
 { {x -> a}, {x -> nextic} }
 )
 ],
 Cases[ p, {l_,eq|eqi,a_,b_,l2_} ->
 {
 q l <= next[l] + l2,
 q l <= next[l] + q eq[a,b]
 }
 ],
 Cases[ p, {l_,neq|neqi,a_,b_,l2_} ->
 {
 q l <= next[l] + l2,
 q l <= l2 + q eq[a,b]
 }
 ],
 Cases[
 DeleteCases[ p,
 {_,halt|goto|jump|goback|eq|eqi|neq|neqi,___}
 ],
 {l_,__} -> q l <= next[l]
 ],
 {
 ic == Plus@@ ((# "("<>ToString[#]<>")")& /@ labels),
 q nextic <= ic,
 ic <= q nextic + qminus1
 }
} // put

(* generate compare equations *)

(
 Cases[ p, {l_,eq|neq,a_,b_,_} ->
 compare[a,b,char[a],char[b]]
 ]
 ~Union~
 Cases[ p, {l_,eqi|neqi,a_,b_,_} ->
 compare[a,b,char[a],b i]
 ]
) /.
 compare[a_,b_,charA_,charB_] ->
{
 {
 eq[a,b] <= i,
 2 eq[a,b] <= ge[a,b] + ge[b,a],
 ge[a,b] + ge[b,a] <= 2 eq[a,b] + i
 },
 {
 geXY <= i,
 256 geXY <= 256 i + charX - charY,
 256 i + charX - charY <= 256 geXY + 255 i
 } /.
 {
 {geXY -> ge[a,b], charX -> charA, charY -> charB},
 {geXY -> ge[b,a], charX -> charB, charY -> charA}
 }
} // put

(* generate auxiliary register equations *)

(* set target t to source s at label l *)
set[t_,s_,l_] :=
 {
 set <= s,
 set <= qminus1 l,
 s <= set + qminus1 (i - l)
 } /.
 set -> set[t,l]

{
 Cases[ p, {l_,set,a_,b_} ->
 set[a,b,l]
 ],
 Cases[ p, {l_,seti,a_,b_} ->
 set[a,b i,l]
 ],
 Cases[ p, {l_,left,a_,b_} ->
 {
 set[a,256a+char[b],l],
 set[b,shift[b],l]
 }
 ],
 Cases[ p, {l_,lefti,a_,b_} ->
 set[a,256a+i b,l]
 ],
 Cases[ p, {l_,right,a_} ->
 set[a,shift[a],l]
 ],
 Cases[ p, {l_,jump,a_,_} :>
 set[a,i "("<>ToString[next[l]]<>")",l]
 ]
} // put

(* generate main register equations *)

defs[r_] := defs[r] = Cases[ p,
 {l_,set|seti|left|lefti|right|jump,r,___} |
 {l_,left,_,r}
 -> l ]

(
 Function[ r,
 {
 r <= qminus1 i,
 r + output q^time ==
 input + q (dontset + Plus@@ (set2[r,#]& /@ defs[r])),
 set == Plus@@ defs[r],
 dontset <= r,
 dontset <= qminus1 (i - set),
 r <= dontset + qminus1 set,
 256 shift <= r,
 256 shift <= i (qminus1 - 255),
 r <= 256 shift + 255 i,
 r == 256 shift + char
 } /. ((# -> #[r])& /@
 {input,output,set,dontset,shift,char}) /.
 set2 -> set
 ] /@ registers
) // put

(* all equations and inequalities are now in eqs; *)
(* start processing *)

eqs = Flatten[eqs]

print@
 "********** combined list of <='s & =='s"
print@
 Short[InputForm@ eqs,10]

(* how many ='s, <='s, registers, labels, variables ? *)

print@StringForm[
 "********** `` =='s, `` <='s, `` total",
 neq = Count[eqs,_==_], nle = Count[eqs,_<=_], Length@ eqs
 ]
print@
 "********** now counting variables"

variables =
 eqs /. Plus|Times|Power|Equal|LessEqual -> List

variables =
 DeleteCases[ Flatten@ variables, _String|_Integer ] // Union

print@StringForm[
"********** `` registers, `` labels, `` variables altogether",
Length@ registers, Length@ labels, nvar = Length@ variables
]
Write[o,variables]

(* convert strings to integers *)

bitmap =
 (FromCharacterCode@ # -> StringJoin[
  Rest@ IntegerDigits[256 + #, 2] /.
  {0 -> "0", 1 -> "1"} ]
 ) & /@ Range[0, 127]

s2i[x_] :=
 ToExpression[
 "2^^" <> StringReplace[StringReverse@ x, bitmap]
 ]

print@
 "********** now converting strings to integers"

eqs = eqs /.
 {eq[x__] -> eq[x], ge[x__] -> ge[x], x_String :> s2i@x}

(* transpose negative terms from rhs to lhs of equation *)

negterms[ (term:(x_Integer _. /; x < 0)) + rest_. ] :=
 term + negterms@ rest

negterms[ _ ] := 0

fix[x_] :=
 (
 x /. l_ == r_ :> l == Expand @ r
 ) /. l_ == r_ :> ( (l - # == r - #)&@ negterms@ r )

(* expand each implication into 7 equations & *)
(* add 9 variables *)

print@
 "********** now expanding <='s"
If[ fulloutput,
 Write[o,OutputForm@
 "********** expand each <="
 ]
]

eqs = eqs /. a_ <= b_ :>
(
 If[ fulloutput, Write[o,a<=b]; Write[o,#]; #, # ]& @
 Module[ {r,s,t,u,v,w,x,y,z},
 {
 fix[r == a],
 fix[s == b],
 t == 2^s,
 (1+t)^s == v t^(r+1) + u t^r + w,
 w + x + 1 == t^r,
 u + y + 1 == t,
 u == 2 z + 1
 }
 ]
)

eqs = Flatten[eqs]

print@
 "********** <='s expanded into =='s"
print@
 Short[InputForm@ eqs,10]
print@
 "********** each <= became 7 =='s and added 9 variables"
print@StringForm[
 "********** so should now have `` =='s and `` variables",
 neq + 7 nle, nvar + 9 nle
 ]
print@StringForm[
 "********** actually there are now `` =='s",
 Length@ eqs
 ]

(* combine all equations into one equation *)

ClearAttributes[ {Plus,Times}, {Orderless,Flat} ]

print@
"********** now combining equations"

eqn =
(
 Plus@@ ( eqs /. l_ == r_ -> (l^2 + r^2) ) ==
 Plus@@ ( eqs /. l_ == r_ -> 2 l r )
)

(***
(* Check that no =='s or <='s have become True or False, *)
(* that no <='s are left, that there are no minus signs, *)
(* and that there is just one == *)
If[ fulloutput,
 trouble[] := (Print@"trouble!"; Abort[]);
 print@
 "********** now checking combined equation";
 eqn /. True :> trouble[];
 eqn /. False :> trouble[];
 eqn /. _<=_ :> trouble[];
 eqn /. x_Integer /; x < 0 :> trouble[];
 eqn[[1]] /. _==_ :> trouble[];
 eqn[[2]] /. _==_ :> trouble[];
]
***)

print@
"********** combined equation"
print@
 Short[InputForm@ eqn,10]
print@StringForm[
 "********** `` terms on lhs, `` terms on rhs",
 Length@ eqn[[1]], Length@ eqn[[2]]
 ]
Write[o,OutputForm@
 "********** combined equation 2"
 ]
Write[o,OutputForm@
 Short[InputForm@ eqn,100]
 ]
Write[o,OutputForm@
 "********** left side"
 ]
Write[o,OutputForm@
 Short[InputForm@ eqn[[1]],50]
 ]
Write[o,OutputForm@
 "********** right side"
 ]
Write[o,OutputForm@
 Short[InputForm@ eqn[[2]],50]
 ]
Write[o,OutputForm@
 "********** first 50 terms"
 ]
Write[o,
 Take[eqn[[1]],+50]
 ]
Write[o,OutputForm@
 "********** last 50 terms"
 ]
Write[o,
 Take[eqn[[2]],-50]]
If[ fulloutput,
 print@
 "********** now writing full equation";
 Write[o,OutputForm@
 "********** combined equation in full"
 ];
 Write[o,
 eqn
 ],
 print@
 "********** now determining size of equation";
 print@StringForm[
 "********** size of equation `` characters",
 StringLength@ ToString@ InputForm@ eqn
 ]
]
print@StringForm[
 "********** elapsed time `` seconds",
 Round[SessionTime[]-t0]
 ]
Print@
 "********** list of =='s left in variable eqs"
Print@
 "********** combined == left in variable eqn"
Print@
 "********** warning: + * noncommutative nonassociative!"
Print@
 "********** (to preserve order of terms & factors in eqn)"
Close@ o
\end{verbatim}
}\chap{lisp.c}{\size\begin{verbatim}
/* high speed LISP interpreter */

#include <stdio.h>

#define SIZE 10000000 /* numbers of nodes of tree storage */
#define LAST_ATOM 128 /* highest integer value of character */
#define nil 128 /* null pointer in tree storage */
#define question -1 /* error pointer in tree storage */
#define exclamation -2 /* error pointer in tree storage */

long hd[SIZE+1], tl[SIZE+1]; /* tree storage */
long vlst[LAST_ATOM+1]; /* bindings of each atom */
long next = LAST_ATOM+1; /* next free cell in tree storage */
long tape; /* Turing machine tapes */
long display; /* display indicators */
long outputs; /* output stacks */
long q; /* for converting expressions to binary */

void initialize_atoms(void); /* initialize atoms */
void clean_env(void); /* clean environment */
void restore_env(void); /* restore dirty environment */
long eval(long e, long d); /* evaluate expression */
/* evaluate list of expressions */
long evalst(long e, long d);
/* bind values of arguments to formal parameters */
void bind(long vars, long args);
long at(long x); /* atomic predicate */
long jn(long x, long y); /* join head to tail */
long eq(long x, long y); /* equal predicate */
long cardinality(long x); /* number of elements in list */
long append(long x, long y); /* append two lists */
/* read one square of Turing machine tape */
long getbit(void);
/* read one character from Turing machine tape */
long getchr(void);
/* read expression from Turing machine tape */
long getexp(long top);
void putchr(long x); /* convert character to binary */
void putexp(long x); /* convert expression to binary */
long out(long x); /* output expression */
void out2(long x); /* really output expression */
long in(); /* input expression */

main() /* lisp main program */
{
 long d = 999999999; /* "infinite" depth limit */
 long v;
 initialize_atoms();
 tape = jn(nil,nil);
 display = jn('Y',nil);
 outputs = jn(nil,nil);
 /* read in expression, evaluate it, & write out value */
 v = eval(in(),d);
 if (v == question) v = '?';
 if (v == exclamation) v = '!';
 out(v);
}

void initialize_atoms(void) /* initialize atoms */
{
 long i;
 for (i = 0; i <= LAST_ATOM; ++i) {
 hd[i] = tl[i] = i; /* so that hd & tl of atom = atom */
 /* initially each atom evaluates to self */
 vlst[i] = jn(i,nil);
 }
}

long jn(long x, long y) /* join two lists */
{
 /* if y is not a list, then jn is x */
 if ( y != nil && at(y) ) return x;

 if (next > SIZE) {
 printf("Storage overflow!\n");
 exit(0);
 }

 hd[next] = x;
 tl[next] = y;

 return next++;
}

long at(long x) /* atom predicate */
{
 return ( x <= LAST_ATOM );
}

long eq(long x, long y) /* equal predicate */
{
 if (x == y) return 1;
 if (at(x)) return 0;
 if (at(y)) return 0;
 if (eq(hd[x],hd[y])) return eq(tl[x],tl[y]);
 return 0;
}

long eval(long e, long d) /* evaluate expression */
{
/*
 e is expression to be evaluated
 d is permitted depth - integer, not pointer to tree storage
*/
 long f, v, args, x, y, z, vars, body;

 /* find current binding of atomic expression */
 if (at(e)) return hd[vlst[e]];

 f = eval(hd[e],d); /* evaluate function */
 e = tl[e]; /* remove function from list of arguments */
 if (f < 0) return f; /* function = error value? */

 if (f == '\'') return hd[e]; /* quote */

 if (f == '/') { /* if then else */
 v = eval(hd[e],d);
 e = tl[e];
 if (v < 0) return v; /* error? */
 if (v == '0') e = tl[e];
 return eval(hd[e],d);
 }

 args = evalst(e,d); /* evaluate list of arguments */
 if (args < 0) return args; /* error? */

 x = hd[args]; /* pick up first argument */
 y = hd[tl[args]]; /* pick up second argument */
 z = hd[tl[tl[args]]]; /* pick up third argument */

 switch (f) {
 case '@': return getbit();
 case '%': return getexp('Y');
 case '#': {v = q = jn(nil,nil); putexp(x); return tl[v];}
 case '+': return hd[x];
 case '-': return tl[x];
 case '.': return (at(x) ? '1' : '0');
 case ',': {hd[outputs] = jn(x,hd[outputs]);
            return (hd[display] == 'Y' ? out(x) : x);
           }
 case '~': return out(x);
 case '=': return (eq(x,y) ? '1' : '0');
 case '*': return jn(x,y);
 case '^': return append((at(x)?nil:x),(at(y)?nil:y));
 }

 if (d == 0) return question; /* depth exceeded -> error! */
 d--; /* decrement depth */

 if (f == '!') {
 clean_env(); /* clean environment */
 v = eval(x,d);
 restore_env(); /* restore unclean environment */
 return v;
 }

 if (f == '?') {
 x = cardinality(x); /* convert s-exp into number */
 clean_env();
 tape = jn(z,tape);
 display = jn('N',display);
 outputs = jn(nil,outputs);
 v = eval(y,(d <= x ? d : x));
 restore_env();
 z = hd[outputs];
 tape = tl[tape];
 display = tl[display];
 outputs = tl[outputs];
 if (v == question) return (d <= x ? question : jn('?',z));
 if (v == exclamation) return jn('!',z);
 return jn(jn(v,nil),z);
 }

 f = tl[f];
 vars = hd[f];
 f = tl[f];
 body = hd[f];

 bind(vars,args);

 v = eval(body,d);

 /* unbind */
 while (at(vars) == 0) {
 if (at(hd[vars]))
 vlst[hd[vars]] = tl[vlst[hd[vars]]];
 vars = tl[vars];
 }

 return v;
}

void clean_env(void) /* clean environment */
{
 long i;
 for (i = 0; i <= LAST_ATOM; ++i)
 vlst[i] = jn(i,vlst[i]); /* clean environment */
}

void restore_env(void) /* restore unclean environment */
{
 long i;
 for (i = 0; i <= LAST_ATOM; ++i)
 vlst[i] = tl[vlst[i]]; /* restore unclean environment */
}

long cardinality(long x) /* number of elements in list */
{
 if (at(x)) return (x == nil ? 0 : 999999999);
 return 1+cardinality(tl[x]);
}

/* bind values of arguments to formal parameters */
void bind(long vars, long args)
{
 if (at(vars)) return;
 bind(tl[vars],tl[args]);
 if (at(hd[vars]))
 vlst[hd[vars]] = jn(hd[args],vlst[hd[vars]]);
}

long evalst(long e, long d) /* evaluate list of expressions */
{
 long x, y;
 if (at(e)) return nil;
 x = eval(hd[e],d);
 if (x < 0) return x; /* error? */
 y = evalst(tl[e],d);
 if (y < 0) return y; /* error? */
 return jn(x,y);
}

long append(long x, long y) /* append two lists */
{
 if (at(x)) return y;
 return jn(hd[x],append(tl[x],y));
}

/* read one square of Turing machine tape */
long getbit(void)
{
 long x;
 if (at(hd[tape])) return exclamation; /* tape finished ! */
 x = hd[hd[tape]];
 hd[tape] = tl[hd[tape]];
 return (x == '0' ? '0' : '1');
}

/* read one character from Turing machine tape */
long getchr(void)
{
 long c, b, i;
 c = 0;
 for (i = 0; i < 7; ++i) {
 b = getbit();
 if (b < 0) return b; /* error? */
 c = c + c + b - '0';
 }
 /* nonprintable ASCII -> ? */
 return (c > 31 && c < 127 ? c : '?');
}

/* read expression from Turing machine tape */
long getexp(long top)
{
 long c = getchr(), first, last, next;
 if (c < 0) return c; /* error? */
 if (top == 'Y' && c == ')') return nil; /* top level only */
 if (c != '(') return c;
 /* list */
 first = last = jn(nil,nil);
 while ((next = getexp('N')) != ')') {
 if ( next < 0 ) return next; /* error? */
 last = tl[last] = jn(next,nil);
 }
 return tl[first];
}

void putchr(long x) /* convert character to binary */
{
 q = tl[q] = jn(( x &  64 ? '1' : '0' ), nil);
 q = tl[q] = jn(( x &  32 ? '1' : '0' ), nil);
 q = tl[q] = jn(( x &  16 ? '1' : '0' ), nil);
 q = tl[q] = jn(( x &   8 ? '1' : '0' ), nil);
 q = tl[q] = jn(( x &   4 ? '1' : '0' ), nil);
 q = tl[q] = jn(( x &   2 ? '1' : '0' ), nil);
 q = tl[q] = jn(( x &   1 ? '1' : '0' ), nil);
}

void putexp(long x) /* convert expression to binary */
{
 if ( at(x) && x != nil ) {putchr(x); return;}
 putchr('(');

 while (at(x) == 0) {
 putexp(hd[x]);
 x = tl[x];
 }

 putchr(')');
}

long out(long x) /* output expression */
{
 out2(x);
 putchar('\n');
 return x;
}

void out2(long x) /* really output expression */
{
 if ( at(x) && x != nil ) {putchar(x); return;}
 putchar('(');

 while (at(x) == 0) {
 out2(hd[x]);
 x = tl[x];
 }

 putchar(')');
}

long in() /* input expression */
{
 long c = getchar(), first, last, next;
 if (c != '(') return c;
 /* list */
 first = last = jn(nil,nil);
 while ((next = in()) != ')')
 last = tl[last] = jn(next,nil);
 return tl[first];
}
\end{verbatim}
}\chap{lisp.rm}{\size\begin{verbatim}
{

(* The LISP Machine! ... *)
(* register machine LISP interpreter *)
(* input in expression, output in value *)

 empty[alist], (* initial association list *)
 set[stack,alist], (* empty stack *)
 set[tape,alist], (* empty tape *)
 set[output,alist], (* empty output *)
 set[depth,"0"], (* no depth limit *)
 set[display,"y"], (* display on *)
 jump[linkreg,eval], (* evaluate expression *)
 neq[value,")",finished],
 set[value,error$value], (* value now ! or ? *)
finished,
 halt[], (* finished ! *)

(* Recursive Return ... *)

return$error,
 set[value,")"],
 goto[unwind],

return0,
 set[value,"0"],
 goto[unwind],

return1,
 set[value,"1"],

unwind,
 pop[linkreg], (* pop return address *)
 goback[linkreg],

(* Recursive Call ... *)

eval,
 push[linkreg], (* push return address *)
 atom[expression,expression$is$atom],
 goto[expression$isnt$atom],

expression$is$atom,
 set[x,alist], (* copy alist *)
alist$search,
 set[value,expression], (* variable not in alist *)
 atom[x,unwind], (* evaluates to self *)
 popl[y,x], (* pick up variable *)
 popl[value,x], (* pick up its value *)
 eq[expression,y,unwind], (* right one ? *)
 goto[alist$search],

expression$isnt$atom, (* expression is not atom *)
 (* split into function & arguments *)
 split[expression,arguments,expression],
 push[arguments], (* push arguments *)
 jump[linkreg,eval], (* evaluate function *)
 pop[arguments], (* pop arguments *)
 eq[value,")",unwind], (* abort ? *)
 set[function,value], (* remember value of function *)

(* Quote ... *)

 neq[function,"'",not$quote],

 (* ' quote *)
 hd[value,arguments], (* return argument "as is" *)
 goto[unwind],

not$quote,

(* If ... *)

 neq[function,"/",not$if$then$else],

 (* / if *)
 popl[expression,arguments], (* pick up "if" clause *)
 push[arguments], (* remember "then" & "else" clauses *)
 jump[linkreg,eval], (* evaluate predicate *)
 pop[arguments], (* pick up "then" & "else" clauses *)
 eq[value,")",unwind], (* abort ? *)
 neq[value,"0",then$clause], (* predicate considered true *)
 (* if not 0 *)
 tl[arguments,arguments], (* if false, skip "then" clause *)
then$clause, (* pick up "then" or "else" clause *)
 hd[expression,arguments],
 jump[linkreg,eval], (* evaluate it *)
 goto[unwind], (* return value "as is" *)

not$if$then$else,

(* Evaluate Arguments ... *)

 push[function],
 jump[linkreg,evalst],
 pop[function],
 eq[value,")",unwind], (* abort ? *)
 set[arguments,value], (* remember argument values *)
 popl[x,value], (* pick up first argument in x *)
 split[y,z,value], (* second argument in y *)
 hd[z,z], (* & third argument in z *)

(* Read Bit ... *)

 neq[function,"@",not$bit],

 (* @ read bit *)
 set[error$value,"!"],
 atom[tape,return$error],
 popl[value,tape],
 eq[value,"0",unwind],
 goto[return1],

not$bit,

(* Read Expression ... *)

 neq[function,"%",not$read],

 (* % read expression *)
 (* keep (reversed) result in expression *)
 (* and keep count of unclosed parens in function *)
 set[expression,"\0"],
 set[function,"\0"],
 (* create constant 8 in z *)
 set[z,"1"], left[z,"1"], left[z,"1"], left[z,"1"],
 left[z,"1"], left[z,"1"], left[z,"1"], left[z,"1"],
main$read$exp$loop, (* repeatedly read 7 bits and convert *)
 (* copy 8 in x *)
 set[x,z], (* loop counter *)
 set[y,"\0"], (* to accumulate all 7 bits *)
 set[error$value,"!"], (* in case reach end of tape *)
read$bit$loop, (* read 7 bits from tape into y *)
 right[x],
 eq[x,"\0",read$bit$loop$finished],
 atom[tape,return$error],
 popl[value,tape],
 eq[value,"0",good$bit],
 eq[value,"1",good$bit],
 set[value,"1"],
good$bit,
 left[y,value],
 goto[read$bit$loop],
(* now 8 in z and reversed bits of char in y *)
read$bit$loop$finished,
 left[x,y],
 neq[y,"\0",read$bit$loop$finished],
 set[y,x],
 (* now bits of char in correct order in y *)
 set[source,printable$ascii$characters],
 set[source2,ascii$bits],
read$char$loop, (* loop through char and bit tables *)
 set[value,"\0"],
 left[value,source], (* may be char we are looking for *)
 (* copy 8 in parens *)
 set[parens,z], (* loop counter *)
 set[x,y], (* copy reversed string of bits in x *)
 set[work,"1"], (* any mismatched bit turns this flag off *)
read$char$loop2, (* loop through each bit of character *)
 right[parens],
 eq[parens,"\0",did$all$bits$match],
 eq[x,source2,bit$matches],
 set[work,"0"],
bit$matches,
 right[x],
 right[source2],
 goto[read$char$loop2], (* check if next bit matches *)
did$all$bits$match,
 eq[work,"0",read$char$loop], (* go to next char *)
 (* found it! *)
 left[expression,value],
 neq[expression,"(",read$exp$skip1],
 left[function,"1"],
read$exp$skip1,
 neq[expression,")",read$exp$skip2],
 right[function],
read$exp$skip2,
 neq[function,"\0",main$read$exp$loop],
 (* now have a complete expression with balanced parens *)
 (* but in reverse order!! *)
 set[value,"\0"],
final$read$exp$loop,
 left[value,expression],
 neq[expression,"\0",final$read$exp$loop],
 neq[value,")",unwind], (* right parens ) all by self *)
 empty[value], (* is changed to () *)
 goto[unwind],

not$read,

(* Convert to Binary ... *)

 neq[function,"#",not$binary],

 (* # convert to binary *)
 set[y,"\0"],
convert$loop, (* look at each char in argument string *)
 set[source,all$ascii$characters],
 set[source2,ascii$bits],
convert$loop2, (* loop through char and bit tables *)
 neq[source,x,convert$loop2$bump], (* find char in
 table of chars, then corresponding place in
 table of bits gives conversion *)
 left[y,source2], left[y,source2], left[y,source2],
 left[y,source2], left[y,source2], left[y,source2],
 left[y,source2],
 goto[convert$loop$bump],
convert$loop2$bump,
 right[source], (* bump table of chars by 1 char *)
                (* bump table of bits  by 7 bits *)
 right[source2], right[source2], right[source2],
 right[source2], right[source2], right[source2],
 right[source2],
 goto[convert$loop2],
convert$loop$bump, (* go to next char in argument string *)
 right[x],
 neq[x,"\0",convert$loop],
 set[value,")"], (* finally add parens & reverse the result *)
convert$loop3,
 left[value,y],
 neq[y,"\0",convert$loop3],
 left[value,"("],
 goto[unwind],

not$binary,

(* Atom & Equal ... *)

 neq[function,".",not$atom],

 (* . atom *)
 atom[x,return1], (* if argument is atomic return true *)
 goto[return0], (* otherwise return false *)

not$atom,

 neq[function,"=",not$equal],

(* = equal *)
compare,
 neq[x,y,return0], (* not equal ! *)
 right[x],
 right[y],
 neq[x,"\0",compare],
 goto[return1], (* equal ! *)

not$equal,

(* Head, Tail & Join ... *)

 split[target,target2,x], (* get head & tail of argument *)
 set[value,target],
 eq[function,"+",unwind], (* + pick head *)
 set[value,target2],
 eq[function,"-",unwind], (* - pick tail *)
 jn[value,x,y], (* * join first argument to second argument *)
 eq[function,"*",unwind],

(* Concatenate ... *)

 neq[function,"^",not$conc],

 (* ^ concatenate *)
 empty[source], (* Will put first argument x into source *)
 set[source2,source], (* & second argument y into source2 *)
 neq[x,"(",conc$skip], (* atomic first argument -> () *)
 set[source,x],
conc$skip,
 neq[y,"(",conc$skip2], (* atomic second argument -> () *)
 set[source2,y],
conc$skip2,
 set[value,source2], (* put second argument in value *)
 right[value], (* & remove initial ( from second argument *)
 set[work,"\0"],
conc$loop1, (* reverse copy first argument into work *)
 left[work,source],
 neq[source,"\0",conc$loop1],
 right[work], (* remove final ) from first argument *)
conc$loop2, (* & reverse work onto front of 2nd argument *)
 left[value,work],
 neq[work,"\0",conc$loop2],
 goto[unwind],

not$conc,

(* Output ... *)

 neq[function,",",not$output],

 (* , output *)
 jn[output,x,output], (* collect output *)
 eq[display,"n",skip$display],
 out[x], (* write argument *)
skip$display,
 set[value,x], (* identity function! *)
 goto[unwind],

not$output,

(* Debug Output ... *)

 neq[function,"~",not$debug],

 (* ~ debug output *)
 out[x], (* write argument *)
 set[value,x], (* identity function! *)
 goto[unwind],

not$debug,

(* Decrement Depth Limit ... *)

 neq[depth,"(",no$limit],
 set[error$value,"?"],
 atom[depth,return$error], (* if limit exceeded, unwind *)
no$limit,
 push[depth], (* push limit before decrementing it *)
 tl[depth,depth], (* decrement it *)

(* Eval ... *)

 neq[function,"!",not$eval],

 (* ! eval *)
 set[expression,x], (* pick up argument *)
 push[alist], (* push alist *)
 empty[alist], (* fresh environment *)
 jump[linkreg,eval], (* evaluate argument again *)
 pop[alist], (* restore old environment *)
 pop[depth], (* restore old depth limit *)
 goto[unwind],

not$eval,

(* Evald ... *)

 neq[function,"?",not$evald],

 (* ? eval depth limited *)
 set[value,x], (* pick up first argument *)
 set[expression,y], (* pick up second argument *)
 (* First argument of ? is in value and *)
 (* second argument of ? is in expression. *)
 (* First argument is new depth limit and *)
 (* second argument is expression to safely eval. *)
 (* Third argument of ? stays in z and *)
 (* is new Turing machine tape. *)
 push[alist], (* save old environment *)
 empty[alist], (* fresh environment *)
 push[output], (* save old output *)
 empty[output], (* empty output *)
 push[tape], (* save old tape *)
 set[tape,z], (* pick up new tape *)
 push[display], (* save old display switch *)
 set[display,"n"], (* turn switch off *)
 (* decide whether old or new depth restriction is stronger *)
 set[x,depth], (* pick up old depth limit *)
 set[y,value], (* pick up new depth limit *)
 neq[x,"(",new$depth], (* no previous limit, *)
 (* so switch to new one *)
 neq[y,"(",old$depth], (* no new limit, *)
 (* so stick with old one *)
choose,
 atom[x,old$depth], (* old limit smaller, so keep it *)
 atom[y,new$depth], (* new limit smaller, so switch *)
 tl[x,x],
 tl[y,y],
 goto[choose],

new$depth, (* new depth limit more restrictive *)
 set[depth,value], (* pick up new depth limit *)
 jump[linkreg,eval], (* evaluate second argument of ? again *)
 pop[display], (* restore display switch *)
 pop[tape], (* restore tape *)
 set[z,output], (* save output for wrapper *)
 pop[output], (* restore output *)
 pop[alist], (* restore environment *)
 pop[depth], (* restore depth limit *)
 neq[value,")",wrap],
error$wrap,
 set[value,error$value], (* convert "no value" to ? or ! *)
 goto[final$wrap],
wrap,
 empty[source2],
 jn[value,value,source2], (* wrap good value in parentheses *)
final$wrap,
 jn[value,value,z], (* & add the outputs *)
 goto[unwind],

old$depth, (* old depth limit more restrictive *)
 jump[linkreg,eval], (* evaluate second argument of ? again *)
 pop[display], (* restore display switch *)
 pop[tape], (* restore tape *)
 set[z,output], (* save output for wrapper *)
 pop[output], (* restore output *)
 pop[alist], (* restore environment *)
 pop[depth], (* restore depth limit *)
 neq[value,")",wrap], (* not error value -> big wrapper *)
 eq[error$value,"?",unwind], (* depth exceeded -> unwind *)
 goto[error$wrap], (* tape finished -> small wrapper *)

not$evald,

(* Defined Function ... *)

 (* bind *)

 tl[function,function], (* throw away & *)
 (* pick up variables from function definition *)
 popl[variables,function],
 push[alist], (* save environment *)
 jump[linkreg,bind], (* new environment *)
 (* (preserves function) *)

 (* evaluate body *)

 hd[expression,function], (* pick up body of function *)
 jump[linkreg,eval], (* evaluate body *)

 (* unbind *)

 pop[alist], (* restore environment *)
 pop[depth], (* restore depth limit *)
 goto[unwind],

(* Evalst ... *)
(* input in arguments, output in value *)

evalst, (* loop to eval arguments *)
 push[linkreg], (* push return address *)
 set[value,arguments], (* null argument list has *)
 atom[arguments,unwind], (* null list of values *)
 popl[expression,arguments], (* pick up next argument *)
 push[arguments], (* push remaining arguments *)
 jump[linkreg,eval], (* evaluate first argument *)
 pop[arguments], (* pop remaining arguments *)
 eq[value,")",unwind], (* abort ? *)
 push[value], (* push value of first argument *)
 jump[linkreg,evalst], (* evaluate remaining arguments *)
 pop[x], (* pop value of first argument *)
 eq[value,")",unwind], (* abort ? *)
 jn[value,x,value], (* add first value to rest *)
 goto[unwind],

(* Bind ... *)
(* input in variables, arguments, alist, output in alist *)

bind, (* must not ruin function *)
 push[linkreg],
 atom[variables,unwind], (* any variables left to bind? *)
 popl[x,variables], (* pick up variable *)
 push[x], (* save it *)
 popl[x,arguments], (* pick up argument value *)
 push[x], (* save it *)
 jump[linkreg,bind],
 pop[x], (* pop value *)
 jn[alist,x,alist], (* (value alist) *)
 pop[x], (* pop variable *)
 jn[alist,x,alist], (* (variable value alist) *)
 goto[unwind],

(* Push & Pop Stack ... *)

push$routine, (* input in source *)
 jn[stack,source,stack], (* stack = join source to stack *)
 goback[linkreg2],

pop$routine, (* output in target *)
 split[target,stack,stack], (* target = head of stack *)
 goback[linkreg2], (* stack = tail of stack *)

(* Split S-exp into Head & Tail ... *)
(* input in source, output in target & target2 *)

split$routine,
 set[target,source], (* is argument atomic ? *)
 set[target2,source], (* if so, its head & its tail *)
 atom[source,split$exit], (* are just the argument itself *)
 set[target,"\0"],
 set[target2,"\0"],

 right[source], (* skip initial ( of source *)
 set[work,"\0"],
 set[parens,"\0"], (* p = 0 *)

copy$hd,
 neq[source,"(",not$lpar], (* if ( *)
 left[parens,"1"], (* then p = p + 1 *)
not$lpar,
 neq[source,")",not$rpar], (* if ) *)
 right[parens], (* then p = p - 1 *)
not$rpar,
 left[work,source], (* copy head of source *)
 eq[parens,"1",copy$hd], (* continue if p not = 0 *)

reverse$hd,
 left[target,work], (* reverse result into target *)
 neq[work,"\0",reverse$hd],

 set[work,"("], (* initial ( of tail *)
copy$tl,
 left[work,source], (* copy tail of source *)
 neq[source,"\0",copy$tl],

reverse$tl,
 left[target2,work], (* reverse result into target2 *)
 neq[work,"\0",reverse$tl],

split$exit,
 goback[linkreg3], (* return *)

(* Join x & y ... *)

jn$routine, (* input in source & source2, *)
 set[target,source], (* output in target *)
 neq[source2,"(",jn$exit], (* is source2 a list ? *)
 set[target,"\0"], (* if not, join is just source1 *)

 set[work,"\0"],
 left[work,source2], (* copy ( at beginning of source2 *)

copy1,
 left[work,source], (* copy source1 *)
 neq[source,"\0",copy1],

copy2,
 left[work,source2], (* copy rest of source2 *)
 neq[source2,"\0",copy2],

reverse,
 left[target,work], (* reverse result *)
 neq[work,"\0",reverse],

jn$exit,
 goback[linkreg3] (* return *)

}
\end{verbatim}
}\end{document}